\documentclass[12pt,preprint]{aastex}
\shorttitle{Close Binaries with Variable Alpha}
\newcommand{\Msun}{\rm M_\odot}
\newcommand{\Rsun}{\rm R_\odot}

\begin{document}

\title{Population Synthesis Studies of Close Binary Systems Using a Variable Common Envelope Efficiency Parameter: I. Dependence upon Secondary Mass}

\author{Michael Politano\altaffilmark{1,2} and Kevin P. Weiler\altaffilmark{3}}

\altaffiltext{1}{Department of Physics, Marquette University, P.O. Box 1881, Milwaukee, WI 53201-1881}

\altaffiltext{2}{Department of Physics and Astronomy, Northwestern University, 2131 Tech Drive, Evanston, IL 60208}

\altaffiltext{3}{Department of Physics, DePaul University, 2219 N. Kenmore Ave., Chicago, IL 60614}

\begin{abstract}

We perform population synthesis calculations of present-day post common envelope binaries (PCEBs) and zero-age cataclysmic variables (ZACVs) using a common envelope efficiency parameter, $\alpha_{CE}$, that is a function of secondary mass, $M_s$.   We investigate three basic possibilities: (1) a standard constant $\alpha_{CE}$ prescription, with $\alpha_{CE}$ = 1.0, 0.6, 0.3, 0.2, 0.1 and 0.05, to provide a baseline for comparison, (2) a power law  dependence, $\alpha_{CE}$ = $(M_s)^n$, with $n$ = 0.5, 1.0 and 2.0, and (3) a dependence in which $\alpha_{CE}$ approaches 1 for large secondary masses and $\alpha_{CE}$ = 0 below some assumed cutoff mass, $\alpha_{CE} = 1-M_{cut}/M_s$, where $M_{cut}$ is the cutoff mass and is equal to 0.0375, 0.075 and 0.15 $\Msun$.  For each population, we compute orbital period, orbital separation, white dwarf mass and secondary mass distributions.

We find that if $\alpha_{CE}$ $\lesssim$ 0.2 in our constant $\alpha_{CE}$ sequence, the predicted present-day ZACV population is significantly modified compared with our standard model ($\alpha_{CE}$ = 1.0).  All prior population synthesis calculations of the formation of CVs only considered values of $\alpha_{CE}$ $\geq$ 0.3 and found that their model populations were not strongly dependent upon the value of $\alpha_{CE}$.   Our results indicate that a much wider range of values for $\alpha_{CE}$, including very low values, must be considered in order for a dependence to be seen.  In our variable $\alpha_{CE}$ sequences for ZACVs, we find that for models in which $\alpha_{CE}$ decreases very rapidly for small secondary masses, the orbital period distribution  below the period gap differs significantly from our standard model.  These differences are most evident in our power law sequence model with $n$ = 2 and in our cutoff mass sequence model with $M_{cut}$ = 0.15 $\Msun$.  In these two models, the fraction of CVs forming with orbital periods below the gap is reduced significantly, the fraction forming in the gap is increased significantly, and both the short-period peak and the minimum period in the ZACV orbital period distribution shift to considerably longer orbital periods compared with our standard model.   We suggest that the observed scarcity of CVs with $P <$ 77 min may possibly provide evidence that progenitor binaries with very low mass secondaries ($M_s \lesssim$ 0.10 $\Msun$) are unable to avoid merger within the common envelope.  We also suggest that if $\alpha_{CE}$ decreases rapidly for small secondary masses, as in our power law sequence with $n \ga$ 1, it is possible that the lower edge of the period gap could be, in part, an imprint of the ZACV population.  Such an imprint could be important in recently-proposed non-standard scenarios of CV secular evolution, such as circumbinary disks, which have difficulty in reproducing the sharpness of the lower edge of the gap.

In our constant $\alpha_{CE}$ model sequence for present-day PCEBs, we find that for all values of $\alpha_{CE}$, the majority of the systems contain secondaries with masses $>$ 0.375 $\Msun$, orbital periods $>$ 1 day and orbital separations $>$ 0.025 AU, with most having periods of $\sim$ 3 days and separations of $\sim$ 0.05 AU.  These models further predict that the present-day population of PCEBs should contain roughly an equal number of systems with He and CO WDs if $\alpha_{CE}$ is globally near unity, but should be clearly dominated by systems containing CO WDs if $\alpha_{CE}$ is globally small ($\lesssim$ 0.30).  In our model sequences of present-day PCEBs in which $\alpha_{CE}$ is a function of secondary mass, the only distribution that varies significantly is the secondary mass distribution.  In the power law model sequence, as $n$ is increased from 0.5 to 2.0, the fraction of PCEBs with $M_s$ $<$ 0.375 $\Msun$ decreases by a factor of $\sim$ 4 from 0.26 to 0.06.  In the $n$ = 2.0 model, there are no present-day PCEBs with secondary masses less than 0.10 $\Msun$.  In the cutoff mass model sequence, significant changes only occur near the cutoff mass and the distributions are nearly identical for $M_s \ga$ 0.375 $\Msun$. 

Based on the results of this investigation, we suggest to theorists who perform detailed hydrodynamical calculations of common envelope evolution that a sequence of models with a fixed mass giant and very low mass secondaries, ranging from 0.3 $\Msun$ down to 0.05 $\Msun$, should be given some priority.  We further suggest to observers that a well-defined, statistically complete sample of PCEBs, particularly with regard to very low mass secondaries, is crucially needed to provide tests of detailed models of common envelope evolution.

\end{abstract}


\section{Introduction}

Common envelope (CE) evolution is believed to play an essential role in the evolution of many close binary systems.  In particular, forming short period ($\sim$ days) close binary systems from wide progenitor binaries ($\sim$ years) necessitates some phase of evolution in which a significant amount of the angular momentum is lost from the system.  In a typical CE scenario, the primary star contacts its Roche lobe during a configuration in which mass transfer to the secondary star is dynamically unstable, such as when the primary is a giant and has a deep convective envelope.  The secondary star cannot accommodate the accretion of the material at such high mass transfer rates, and instead becomes quickly engulfed in the envelope of the primary (see \citealp{ibe93,taa00} for recent reviews).  The secondary and the core of the primary star are now orbiting within the envelope of the primary, which has become "common" to both components and which is not co-rotating with respect to the orbit of these components.  Dynamical friction between the orbiting components and the non-corotating envelope cause a dissipation of the orbital energy of the components into the CE.  If enough energy is able to be transferred from the orbit of the engulfed components to the envelope before the components merge, then the envelope may be able to become unbound and be shed.  The result is a much shorter period binary consisting of the core of the primary and the original secondary, which exits the CE relatively unscathed (e.g., \citealp{web88, hje91}). 

Modeling of the CE phase has been undertaken by several authors and dramatic improvements in sophistication have been made over the past thirty years \citetext{e.g., \citealp{mey79}; a series of 11 papers by R. Taam, P. Bodenheimer and co-workers, the latest being \citealp{san00}; \citealp{ras96}}.  The most recent papers on this topic involve 3-dimensional hydrodynamic calculations of this phase of evolution (e.g., \citealp{ras96,san98,san00,dM03a,dM03b}). Despite these advances, the latest model calculations are not yet able to follow the evolution through full envelope ejection and thus can only place upper limits on the final (post-CE) orbital separations (see \citealp{san98,san00}).  This inability is primarily due to the large range in spatial scales that need to be resolved in following the evolution from start (orbital separation of $\sim$ 100-1000 $\Rsun$) to finish (orbital separation of $\sim$ a few $\Rsun$).  Consequently, the results of these detailed model calculations of the CE phase have not been able to be included into population synthesis calculations as yet. 

Instead, population synthesis calculations have relied on simple energy considerations to relate the post- and pre-CE orbital parameters \citep{tut79}.  Although the exact prescription varies somewhat from author to author, a typical expression used in population synthesis calculations is given below (e.g., \citealp{ibe85,pol88,pol96,dis94,wil05}): 
\begin{equation}
  -\alpha_{CE}\left(\frac{GM_cM_s}{2a_f} - \frac{GM_pM_s}{2a_i}\right)
     = -\frac{GM_e(M_e + M_c)}{\lambda R_i}
\end{equation}
\noindent where $M_p$ is the total primary mass at the onset of the CE phase, $M_s$ is the mass of the secondary, $M_c$ is the core mass of the primary, $M_e$ is the envelope mass of the primary ($M_e$ = $M_p-M_c$), $R_i$ is the radius of the primary at the onset of the CE phase (and is therefore equal to the radius of the primary's Roche lobe), $a_i$ and $a_f$ are the pre-and post-CE orbital separations, respectively, $\lambda$ is a dimensionless factor related to the structure of the giant star, and $\alpha_{CE}$ is a parameter that measures the efficiency at which orbital energy is able to be used in unbinding the CE.  Essentially, this prescription approximates the CE phase by saying that when some fraction of the orbital energy released in the spiral-in process equals the initial gravitational binding energy\footnote{We note that some authors include the thermodynamic internal energy of the envelope on the right side in eq. 1 (e.g., \citealp{han95a,pod03}).  The right side of eq. 1 would then represent the total binding energy of the envelope (gravitational potential energy + internal energy).  Nevertheless, the meaning of $\alpha_{CE}$ as a parameter that measures the efficiency of energy transfer from the orbit to the envelope remains the same regardless of whether the total binding energy or the gravitational binding energy is used (see \citealp{han95a}).} of the primary's envelope, the CE becomes unbound and is presumably (somehow) ejected. The parameter, $\alpha_{CE}$, embodies a major uncertainty in this simplified prescription. 

Standard formation models of CVs have typically chosen $\alpha_{CE}$ = 1.0 and/or $\alpha_{CE}$ = 0.3 (e.g., \citealp{pol88,pol96,dek92,han95a,how01,wil05}).  The latter choice reflects the few estimates of $\alpha_{CE}$ from detailed models of the CE phase.  These estimates suggest that $\alpha_{CE}$ is less than 1.0 and may be closer to 0.3 to 0.5, depending on the component masses of the binary (\citealp{taa96,san98,san00}; also see Fig. 1).  Regardless of the exact value chosen for it, virtually all population synthesis calculations have assumed that $\alpha_{CE}$ is a global constant for all binaries in the population.  However, it is unlikely that this is indeed the case.  A more realistic assumption is that the value of $\alpha_{CE}$ depends on the orbital parameters of the binary, such as the mass of the secondary (see \citealp{pol04}) and the state of evolution of the giant (e.g., its core mass and total mass).    In this series of papers, therefore, we undertake the first systematic investigation of the effect of a variable $\alpha_{CE}$ (i.e., one that depends on the orbital parameters of the binary) on population synthesis of close binaries.  Specifically, in this first paper, we compute population models of present-day post-CE binaries (PCEBs) and zero-age cataclysmic variables (ZACVs) using a CE efficiency parameter that depends upon the mass of the secondary star.  In future papers in this series, we will investigate separately the dependence of $\alpha_{CE}$ on other orbital parameters.  

A central goal in undertaking the present study is to provide guidance to researchers who calculate detailed hydrodynamical models of CE evolution.  Such models are highly computationally intensive (Taam, priv. comm. 2005).  Population synthesis calculations, on the other hand, are relatively quick and are ideally suited for identifying interesting regimes within a specified parameter space that bear further, detailed investigation.  In addition, identification of these regimes may suggest future observations to provide tests of the hydrodynamical models.  

Further motivation for the present investigation comes from recent work by \citet{pol04}, who investigated the formation of cataclysmic variables (CVs) with brown dwarf (BD) secondaries.  For standard assumptions (i.e., $\alpha_{CE}$ = 1 and a flat mass ratio distribution in zero-age main sequence [ZAMS] binaries), his models predict that 18\% of the present-day population of ZACVs contain substellar secondaries.    Unlike period-bounce CVs, where the secondary became substellar as a result of mass transfer to the WD, the secondaries in Politano's study were born substellar.  What is particularly interesting about ZACVs with substellar secondaries is that the majority of them (80\%) have orbital periods shorter than 77 min \citep{pol04}, the observed orbital period minimum in CVs.  From his models, Politano predicts that as much as 15\% of the present-day CV population could contain secondaries that were born substellar and have orbital periods less than 77 min.  Yet of the $\sim$ 600 CVs with known orbital periods \citep{rit03}, only three have orbital periods less than 77 min: V485 Cen (59 min), EI Psc (64 min) and J1507+5230 (67 min).  Moreover, it is highly unlikely that the secondaries in V485 Cen and EI Psc are BDs \citep{aug96, tho02, uem02}.  \citet{pol04} suggests that one possible reason for this large discrepancy between his models and observations is that progenitor binaries that contain BD secondaries may not be able to survive the CE phase.  This would imply that $\alpha_{CE}$ is a function of secondary mass and that there exists a cutoff secondary mass below which merger is inescapable.  This is not unreasonable, since the rate of dissipation of orbital energy within the CE will decrease as the secondary mass decreases.  Consequently, below some secondary mass, this dissipation will be insufficient to drive hydrodynamical motion but rather will be carried away by convection within the CE (R. Taam, 2003, private communication).   We examine the implications of Politano's suggestion in this paper by considering a functional dependence of $\alpha_{CE}$ on $M_s$ that contains a low-mass cutoff.

In the next section, we describe the method used in our population synthesis calculations and the functional relationships for $\alpha_{CE}$($M_s$) that we investigated.  In section 3, we present our model results for the present-day populations of PCEBs and ZACVs.  In section 4, we discuss our results and  suggest future model calculations of the CE phase and future observations of PCEBs.  We close with a summary of our main conclusions.

\section{Method}

A Monte Carlo population synthesis code has been used in this study.  This code is the same as the one used in \citet{pw06} and the physics is described in \citet{pol88,pol96,pol04} except as noted below.  We refer the reader to those papers for detailed discussions, and here summarize only the key assumptions and features of the code and the modified treatment of the CE phase. 

In our Monte Carlo calculations, we begin with 10$^7$ ZAMS binaries.  Following a standard approach, we assume that the distribution of ZAMS binaries is the product of separate distributions over primary mass, mass ratio and orbital period.  We use a \citet{mil79} IMF for the primary masses, which we reproduce numerically using the following Monte Carlo representation \citep{egg89}:
\begin{equation}
   M_p(u) = 0.19u[(1 - u)^{3/4} + 0.032(1 - u)^{1/4}]^{-1} ,
\end{equation}
where $u$ is a random number uniformly distributed between 0 and 1.  We assume a ZAMS orbital period distribution that is flat in log $P$ \citep{abt83}\footnote{We note that such a distribution is inconsistent with the orbital period distributions found by \citet{gri85} and \citet{duq91}.  However, the use of a ZAMS period distribution that is flat in log P (or an orbital separation  distribution that is flat in log a) is widespread in the literature (e.g., \citealp{hur02, wil05, zha05}) and thus facilitates comparison.  Additionally, model populations of ZACVs and PCEBs that we calculated using a Griffin or Duquennoy and Mayor-type period distribution produced substantially the same conclusions as those calculated using a period distribution that is flat in log P.} and a ZAMS mass ratio distribution that is flat in $q$ (i.e., $g(q)\,dq = 1\,dq$), where $q$ = $M_s$/$M_p$ \citep{duq91,maz92,gol03}.  Once the primary mass and mass ratio have been generated for a given binary, the secondary mass is determined by their product:  $M_s$ = $q\,M_p$.  The maximum primary mass in our calculations is 10 $\Msun$.  The minimum secondary mass that we consider is 0.013 $\Msun$.   

To form a ZACV in our population synthesis code, a given ZAMS binary is evolved to the point when the primary contacts its Roche lobe while it is ascending the giant branch(es).  Wind loss during ascent of the giant branch(es) is incorporated via a \citet{rei75} prescription.  Relationships between the radius of the giant, its core mass and its total mass are derived from analytic fits to detailed stellar evolution models (see \citealp{pol88,pol96} for an in-depth discussion and for detailed references).   Simple energetic considerations are still used to relate the pre- and post-CE orbital separations for the CE phase (see eq. 1).  Consistent with other studies (e.g., \citealp{wil05}), we choose $\lambda$ = 1.  However, we now allow $\alpha_{CE}$ to be a function of secondary mass (see below).  Since the CE phase is so brief (e.g., \citealp{ibe93,taa00}), the mass of the core remains essentially unchanged during the CE phase and, thus, the white dwarf (WD) mass upon exit of the CE phase is equal to the core mass of the giant at the onset of the CE phase.  Lastly, the classical prescription for gravitational radiation \citep{lan51} and the prescription for disrupted magnetic braking ($\gamma$ = 2) from \citet{rap83} are used to describe angular momentum losses during the post-CE phase.  We use detailed stellar models from the Lyon group \citep{cb97,bar98,bar03,cha00} for secondaries with masses less than $\sim$ 0.5 $\Msun$, including substellar secondaries.  For secondaries with masses greater than $\sim$ 0.5 $\Msun$, we use fits to stellar models from Webbink (see \citealp{pol88,pol96}).  In our models, the secondary is fully convective for $M_s\,\leq\,$0.37 $\Msun$.  

In addition to ZACVs, we model the present-day population of PCEBs.  To avoid ambiguity, we note that when we use the term ``PCEBs'' in this paper, we are specifically referring to binaries that (1) are detached, (2) contain a WD primary and a MS or BD secondary, and (3) have undergone a single CE phase.  Clearly, other important types of close binaries exist that have undergone a single CE phase--for example, systems containing a giant secondary and a WD, such as symbiotic stars or Ba stars (e.g., \citealp{yun95,han95b}).  However, we have limited the scope of our calculations to only PCEBs that satisfy the above three criteria.

The same code as described above for ZACVs was used for PCEBs, but with two important modifications: 1) the secondary radius must be smaller than its Roche lobe radius at the present epoch and 2) we do not impose an upper limit on the binary's mass ratio in order to satisfy stability against thermal or dynamical timescale mass transfer (see \citealp{pol96}), since the system remains detached at the present epoch.  In calculating the population of present-day PCEBs, four timescales are of interest: (1) $t_b$, the time that the progenitor binary was formed (measured from the beginning of the Galaxy), (2) $t_{ev,p}$, the time it takes the primary to evolve off of the main sequence, become a giant, and contact its Roche lobe to initiate the CE phase, (3) $t_{PCEB}$, the time from the end of the CE phase until the present epoch, and (4), $t_{Gal}$, the age of the Galaxy.  These four timescales must satisfy the following constraint,
$t_{b} + t_{ev,p} + t_{PCEB} =  t_{Gal}$.  We make the following assumptions regarding these time scales: (1) the stellar birth rate throughout the Galaxy's history has been constant, (2) the CE phase is so rapid that the time spent in it by the binary is negligible compared to the other time scales (e.g., \citealp{ibe93,taa00}), and (3) the age of the Galaxy is 10$^{10}$ yrs. 

For both populations, we have now allowed $\alpha_{CE}$ to be a function of secondary mass.  We do not change the formal structure of eq. 1, but simply replace $\alpha_{CE}$ by a specified function of secondary mass.  There are only a handful of values of $\alpha_{CE}$ that have been determined from detailed hydrodynamical calculations of the CE phase \citep{san98,san00}, and even these are only upper limits since Sandquist et al. were not able to follow the evolution completely through envelope ejection because of spatial resolution problems.  Consequently, detailed models of the CE phase are able to provide little guidance in constructing an assumed dependence of $\alpha_{CE}$ on $M_s$.   We therefore chose two simple functional forms:  (1) a power law dependence, $\alpha_{CE}$ = $(M_s)^n$, with $n$ = 0.5, 1.0, and 2.0 and $M_s$ in solar masses, and (2) a dependence in which $\alpha_{CE}$ approaches 1 for large secondary masses and is zero below some assumed cutoff mass, $\alpha_{CE} = 1-M_{cut}/M_s$, where $M_{cut}$ is the cutoff mass.  We set $M_{cut}$ = 0.0375, 0.075, and 0.15$\,\Msun$.  These cutoff masses represent 1/2, 1, and 2 times the substellar transition mass, repectively.  Figure 1 shows the $\alpha_{CE}$ estimates from detailed hydrodynamical models (x's in the figure) along with our two assumed functional relationships for $\alpha_{CE}$($M_s$).   As seen in the figure, our assumed relationships span the parameter space containing the estimates of $\alpha_{CE}$ from the detailed models.  

In addition to the variable $\alpha_{CE}$ models, we also calculated constant $\alpha_{CE}$ models for six values of $\alpha_{CE}$:  1.0, 0.6, 0.3, 0.2, 0.1, and 0.05.  These constant $\alpha_{CE}$ models provide a baseline for comparison.  In this paper, we shall refer to the $\alpha_{CE}$ = 1.0 model as our ``standard'' model.

To avoid adding further uncertainty to our investigation, we have not normalized our model distributions to provide absolute numbers of PCEBs or ZACVs.  Instead, our results will be described comparatively and will typically fall into one of two categories:  a comparison of relative features within a given model distribution or a comparison of a particular feature across a given model sequence or across all model sequences.

\section{Results}

In this section, we present our model populations for present-day PCEBs and ZACVs calculated using the $\alpha_{CE}$($M_s$) relationships described in the previous section. For each model population, four distributions were calculated:  orbital period, orbital separation, secondary mass, and WD mass.  Our model populations of present-day PCEBs are discussed in section 3.1 and our model populations of present-day ZACVs are discussed in section 3.2.  Selected results for each model are given in Table 2 for PCEBs and in Table 3 for ZACVs.  In these tables, the constant $\alpha_{CE}$ models are designated by ``CON'' and the value of $\alpha_{CE}$, the power law models are designated by ``PL'' and the value of the exponent $n$, and the cutoff mass models are designated by ``CUTM'' and the value of $M_{cut}$. 

Before proceeding to the specific populations, we discuss the fraction of systems that survive the CE phase, $f_{surv}$, for each model.  We define this fraction as the number of systems that escape merger during the CE phase divided by the total number of systems that entered the CE phase.  We note that for our initial population of 10$^7$ binaries, $\sim$ 17\% of them enter a CE phase at some point during the Galaxy's history due to the primary filling its Roche lobe as a giant.  
Values of $f_{surv}$ are listed in Table 1.   For the CON model sequence, we find that $f_{surv}$ decreases roughly as $(\alpha_{CE})^{0.5}$ from 0.50 to 0.07 as $\alpha_{CE}$ is decreased from 1.0 to 0.05.  Interestingly, the fraction of systems that survive CE evolution does not vary significantly in either the PL or the CUTM model sequences.  In fact, if the CON 1.0 model is included as $n$ = 0 in the PL sequence, then over a wide range of dependences from $\alpha_{CE}$ independent of $M_s$ ($n$ = 0) to $\alpha_{CE}$ very strongly dependent on $M_s$ ($n$ = 2), we find that $f_{surv}$ decreases only slightly from 0.50 to 0.39.  For the CUTM model sequence, a similar trend is found, with $f_{surv}$ decreasing from 0.48 to 0.43 as $M_{cut}$ is increased from 0.0375 to 0.15$\,\Msun$.

\subsection {PCEBs}

\subsubsection {$\alpha_{CE}$ = constant}

The distributions of the orbital periods (top panels) and the orbital separations (bottom panels) in present-day PCEBs for $\alpha_{CE}$ = 1.0, 0.6, and 0.3 (left panels) and for $\alpha_{CE}$ = 0.2, 0.1, and 0.05 (right panels) are shown in Figure 2.  The distributions of the secondary masses (top panels) and the WD masses (bottom panels) in present-day PCEBs for the same choices of $\alpha_{CE}$ are shown in Figure 3.   

Qualitatively, we find that decreasing $\alpha_{CE}$ from 1.0 to 0.05 yields little change in the overall shapes of the orbital period, orbital separation and secondary mass distributions for PCEBs.  As discussed in \citet{pw06}, the sharp drop in the number of present-day PCEBs with secondary masses greater than 0.37 $\Msun$ is due to the increased efficiency of magnetic braking compared with gravitational radiation in shrinking the orbit and bringing systems into contact by the present epoch.  Consequently, only PCEBs that exit the CE phase with relatively wide orbits (i.e., $a \ga$ 0.025 AU or $P \ga$ 1 day) remain detached at the present epoch.  This is reflected in the large increase in the number of PCEBs near $P$ = 1 day and $a$ = 0.025 AU for all values of $\alpha_{CE}$ in Fig. 2.  

For the purpose of comparison and to expedite the discussion, throughout this paper we will denote PCEB secondaries with M$_s < $ 0.375 $\Msun$ as ``low-mass'' secondaries and those with M$_s >$ 0.375 $\Msun$ as ``high-mass'' secondaries (the choice of 0.375 rather than 0.37 has to due with how the distribution was binned).   Similarly, throughout this paper we will denote present-day PCEBs with $P <$ 1 day or $a <$  0.025 AU as ``short-period'' or ``close'' PCEBs and those with $P >$ 1 day or $a >$ 0.025 AU as ``long-period'' or ``wide'' PCEBs.

Quantitatively, we find that as $\alpha_{CE}$ is decreased from 1.0 to 0.05, the fraction of present-day PCEBs with low-mass secondaries decreases slightly from $\sim$ 1/3 to 1/5.  The fraction of present-day short-period PCEBs varies slightly, increasing from 0.23 for $\alpha_{CE}$ = 1.0 to a maximum of 0.27 and then decreasing to 0.19 for $\alpha_{CE}$ = 0.05.  For all values of $\alpha_{CE}$, the orbital period distribution peaks at $\sim$ 3 days and the orbital separation distribution peaks at $\sim$ 0.05 AU.  Thus, over a wide range of values for $\alpha_{CE}$ in our CON sequence, our models predict that the majority of present-day PCEBs should contain high-mass secondaries, have long orbital periods and wide orbits, with most having periods of $\sim$ 3 days and separations of $\sim$ 0.05 AU.

The distribution that is most affected by decreasing $\alpha_{CE}$ is the WD mass distribution.  As $\alpha_{CE}$ is decreased from 1.0 to 0.05, the fraction of He WDs ($M_{WD} \lesssim$ 0.5 $\Msun$) in PCEBs decreases by a factor of $\sim$ 20, from 0.51 to 0.02, and the fraction of CO WDs\footnote{Our code does not separately calculate the formation of PCEBs or ZACVs with ONeMg WDs.  Instead, we assume that some fraction of CVs with high mass WDs contain ONeMg WDs.} ($M_{WD} \ga$ 0.5 $\Msun$) increases correspondingly from 0.49 to 0.98.  Therefore, our CON model sequence predicts that the present-day population of PCEBs should contain roughly an equal number of systems with He and CO WDs if $\alpha_{CE}$ is globally near unity, but should be clearly dominated by systems containing CO WDs if $\alpha_{CE}$ is globally small ($\lesssim$ 0.30).

Lastly, we find that the predicted relative number of present-day PCEBs compared with our standard model decreases monotonically as $\alpha_{CE}$ is decreased in our CON model sequence.  For $\alpha_{CE}$ = 0.05, the number of PCEBs is approximately an order of magnitude smaller than in our standard model.

\subsubsection {$\alpha_{CE}$ = $(M_s)^n$}

The distributions of the orbital periods (top left), the orbital separations (top right), the secondary masses (bottom left) and the WD masses (bottom right) in PCEBs at the present epoch for $\alpha_{CE}$ = $(M_s)^n$, where $n$ = 0.5, 1.0, and 2.0, are shown in Figure 4.

Qualitatively, we find little change in the overall shapes of the distributions as $n$ is varied, except for the secondary mass distribution.  Quantitatively, the dominance of long-period systems is even stronger in the PL model sequence than in the CON model sequence.  The fraction of short-period PCEBs is 0.19 for $n$ = 0.5 and decreases to 0.08 for $n$ = 2.0.  As in the CON models, the distributions over orbital period and separation peak at $\sim$ 3 days and $\sim$ 0.05 AU, respectively.  The WD mass distribution is not affected as strongly in the PL models compared with the CON models.  As $n$ is increased from 0.5 to 2.0, the fraction of He WDs decreases slightly from 0.45 to 0.34 and the fraction of CO WDs correspondingly increases from 0.55 to 0.66.  Thus, in our PL model sequence of present-day PCEBs, the majority of the systems have long orbital periods, wide orbits and contain CO WDs regardless of whether the dependence of $\alpha_{CE}$ on $M_s$ is weak or strong.

The secondary mass distribution is strongly affected by the choice of $n$.  As $n$ is increased from 0.5 to 2.0, the fraction of PCEBs with low-mass secondaries decreases by a factor of $\sim$ 4, from 0.26 to 0.06.  The decrease in the number of low-mass secondaries is particularly severe for the $n$ = 2.0 model.  In this case, there are no present-day PCEBs with secondary masses less than 0.10 $\Msun$.   

The predicted relative number of present-day PCEBs compared with our standard model is 0.80 for $n$ = 0.5 and decreases to 0.50 for $n$ = 2.0.

\subsubsection {$\alpha_{CE} = 1-M_{cut}/M_s$}

The distributions of the orbital periods (top left), the orbital separations (top right), the secondary masses (bottom left) and the WD masses (bottom right) in PCEBs at the present epoch for $\alpha_{CE} = 1-M_{cut}/M_s$, where $M_{cut}$ = 0.0375$\,\Msun$, 0.075$\,\Msun$, and 0.15$\,\Msun$, are shown in Figure 5.  

The orbital period, orbital separation, and WD mass distributions are not affected significantly as the cutoff mass is varied.  Inspection of Table 2 shows that the fraction of systems with He WDs, the fraction with short orbital periods and the fraction with small orbital separations each change only slightly as the cutoff mass is increased from 0.0375 to 0.15 $\Msun$.  Further, the values of these fractions are not very different from those in our standard model.  

The secondary mass distribution is affected, but the effect is much more localized than it was for the PL model sequence.  Significant changes occur only for low-mass secondaries and the distributions are nearly identical above $\sim$ 0.375 $\Msun$.  The fraction of present-day PCEBs with low-mass secondaries decreases from 0.33 to 0.21 as the cutoff mass is increased.    For $M_{cut}$ = 0.0375 $\Msun$, the fraction with substellar secondaries is 0.007.  This is about a factor of 4 smaller than in our standard model.  The fraction with substellar secondaries is obviously zero for $M_{cut}$ = 0.075 $\Msun$ or 0.15 $\Msun$. 

We find that the predicted relative number of present-day PCEBs compared with our standard model decreases modestly from 0.92 to 0.72 as the cutoff mass is increased from 0.0375 to 0.15 $\Msun$.

\subsection {ZACVs}

\subsubsection {$\alpha_{CE}$ = constant}

The distributions of the orbital periods (top panels) and the orbital separations (bottom panels) in ZACVs at the present epoch for $\alpha_{CE}$ = 1.0, 0.6, and 0.3 (left panels) and for $\alpha_{CE}$ = 0.2, 0.1, and 0.05 (right panels) are shown in Figure 6.  The distributions of the secondary masses (top panels) and the WD masses (bottom panels) in ZACVs at the present epoch for the same choices of $\alpha_{CE}$ are shown in Figure 7.   

Varying $\alpha_{CE}$ from 1.0 to 0.3 (left panels in Figures 6 and 7) in our CON model sequence has qualitatively little effect on any of the distributions.  The fraction of systems containing He WDs decreases slightly from 0.58 to 0.46 and there is a slight increase in the fraction of systems with orbital periods above 2.25 hr (in and above the period gap).  These results are in agreement with previous population synthesis calculations of ZACVs that chose $\alpha_{CE}$ = 1.0 and $\alpha_{CE}$ = 0.3 (e.g., \citealp{dek92}).  

However, all distributions show marked differences as $\alpha_{CE}$ is varied from 0.2 to 0.05 (right panels in Figures 6 and 7).  From Table 3, the predicted fraction of CVs formed:  below the period minimum (P $<$ 77 min) decreases by a factor of 3 from 0.15 to 0.05;  below the period gap (P $<$ 2.25 hr) decreases by a factor of 2 from 0.53 to 0.27;  in the period gap (2.25 hr $<$ P $<$ 2.75 hr) increases from 0.11 to 0.16, and above the period gap (P $>$ 2.75 hr) increases from 0.35 to 0.56.  The predicted fraction of CVs formed with low-mass secondaries decreases from $\sim$ 2/3 to 1/2.  The predicted fraction of CVs forming with He WDs decreases by greater than a factor of 3 from 0.41 to 0.12 and, correspondingly, the fraction of CVs forming with CO WDs increases from 0.59 to 0.88.  

Therefore, in our CON model sequence of present-day ZACVs, if $\alpha_{CE}$ $\ga$ 0.30, the majority of the systems contain He WDs and have orbital periods below the period gap.  On the other hand, if $\alpha_{CE}$ $\lesssim$ 0.20, then the majority of the systems contain CO WDs and there is a significant shift in the orbital period distribution towards periods in and above the gap.  The majority of ZACVs contain low-mass secondaries regardless of the value of $\alpha_{CE}$.

Lastly, we find that the predicted number of present-day ZACVs compared with our standard model decreases monotonically as $\alpha_{CE}$ is decreased in our CON model sequence.  For $\alpha_{CE}$ = 0.05, the number of ZACVs is approximately a factor of 4 smaller than in our standard model.

\subsubsection {$\alpha_{CE}$ = $(M_s)^n$}

The distributions of the orbital periods (top left), the orbital separations (top right), the secondary masses (bottom left) and the WD masses (bottom right) in ZACVs at the present epoch for $\alpha_{CE}$ = $(M_s)^n$, where n = 0.5, 1.0, and 2.0, are shown in Figure 8.

Unlike PCEBs, all ZACV distributions are affected significantly by the choice of $n$, and these effects are not restricted solely to $n$ = 2.  In Table 3, we see that as $n$ is increased from 0.5 to 2.0, the fraction of CVs formed with orbital periods: less than 77 min is reduced from 0.15 to zero;  below the period gap decreases by a factor of $\sim$ 7 from 0.58 to 0.09;  in the period gap increases by almost a factor of 3 from 0.07 to 0.18;  and above the period gap doubles from 0.35 to 0.72.  The fraction of CVs formed with low-mass secondaries decreases by almost a factor of 2 from 0.67 to 0.35.   For $n$ = 2, no CVs are formed with secondary masses less than $\sim$ 0.10 $\Msun$.  The fraction of CVs forming with He WDs decreases by almost a factor of 6 from 0.46 to 0.08 as $n$ is increased from 0.5 to 2.0 and, correspondingly, the fraction of CVs forming with CO WDs increases from 0.54 to 0.92.  

Thus, in our PL model sequence of present-day ZACVs, the majority of the systems contain He WDs regardless of whether the dependence of $\alpha_{CE}$ on $M_s$ is weak ($n$ = 0.5) or strong ($n$ = 2.0).  Systems with low-mass secondaries dominate if the dependence of $\alpha_{CE}$ on $M_s$ is weak to moderate ($n \leq$ 1), whereas systems with high-mass secondaries dominate if the dependence is strong.  Finally, systems with orbital periods below the period gap are favored if the dependence of $\alpha_{CE}$ on $M_s$ is weak, while systems in and above the period gap are favored if the dependence is moderate to strong.  

The relative number of ZACVs compared with our standard model is 0.84 for $n$ = 0.5 and decreases to 0.37 for $n$ = 2.0.

\subsubsection {$\alpha_{CE} = 1-M_{cut}/M_s$}

The distributions of the orbital periods (top left), the orbital separations (top right), the secondary masses (bottom left) and the WD masses (bottom right) in ZACVs at the present epoch for $\alpha_{CE} = 1-M_{cut}/M_s$, where $M_{cut}$ = 0.0375, 0.075, and 0.15 $\Msun$, are shown in Figure 9.  

Again, unlike PCEBs, all ZACV distributions are affected by the choice of $M_{cut}$, although the effects are significant only for the lower end of each distribution (i.e., short orbital periods/separations and low-mass secondaries/WDs).  We find that as the cutoff mass is increased from 0.0375 $\Msun$ to 0.15 $\Msun$ in our CUTM sequence, the fraction of CVs formed with orbital periods: less than 77 min decreases from 0.18 to zero;  below the period gap decreases by a factor of 2 from 0.62 to 0.31;  in the period gap nearly triples from 0.05 to 0.13;  and above the period gap increases from 0.34 to 0.55.  The fraction of CVs formed with low-mass secondaries decreases from 0.68 to 0.48.  This decrease is similar to the decrease in the fraction of low-mass secondaries observed in the PL model sequence.  The overall number of CVs forming with He WDs decreases, while the overall number of CVs forming with CO WDs remains approximately the same (see Fig. 9), causing the fraction of present-day ZACVs with He WDs to decrease by almost a factor of 2 from 0.53 to 0.28 and the fraction with CO WDs to increase correspondingly from 0.47 to 0.72.  

Thus, in our CUTM model sequence of present-day ZACVs, the majority of the systems contain low-mass secondaries and have orbital periods below the period gap if $M_{cut} \leq$ 0.075 $\Msun$, while the majority of the systems contain high-mass secondaries and have orbital periods above the gap if $M_{cut}$ = 0.15 $\Msun$.  Further, approximately the same number of systems contain He WDs and CO WDs if $M_{cut} \leq$ 0.075 $\Msun$, while the majority of the systems contain CO WDs if $M_{cut}$ = 0.15 $\Msun$.

Because the secondary star fills its Roche lobe in a ZACV, unlike in a PCEB, there is a strong correlation between the mass of the secondary and the orbital period in a ZACV.  Consequently, the existence of a cutoff mass for the secondary implies the existence of a corresponding minimum orbital period in the ZACV orbital period distribution.  For $M_{cut}$ = 0.0375 $\Msun$, there are no systems with the orbital periods shorter than 43 min.  This minimum period is the same as in our standard model.  However, when $M_{cut}$ is increased to 0.075 $\Msun$, the minimum period increases to 55 min (see Fig. 9).  Finally, when $M_{cut}$ is increased to 0.15 $\Msun$, the cutoff period increases to 102 min, over twice the value in our standard model.  These results are significant since they clearly imply that if a cutoff mass for the secondary indeed exists and is greater than the substellar transition mass, the standard model for the formation of CVs below the period gap will need to be modified (see section 4.2). 

Lastly, we find that the relative number of ZACVs compared with our standard model decreases from 0.86 to 0.52 as the cutoff mass is increased from 0.0375 to 0.15 $\Msun$.

\section{Discussion}

\subsection {Observations of PCEBs: the current state}

\subsubsection {Large Surveys}

The observed number of detached WD-MS systems continues to increase dramatically.  As few as six years ago, there were only $\sim$ 40 known (Hillwig et al. 2000).  That number has risen to over 800, mainly due to searches for them in large observational surveys such as the Sloan Digital Sky Survey (SDSS) and the Two Micron All Sky Survey (2MASS).  For example, \citet{sil06} have recently compiled a catalog of 746 spectroscopically-identified, detached close binary systems in the SDSS through the Fourth Data Release \citep{ade06}.  The vast majority of these systems ($\sim$ 700) are detached WD-M dwarf binaries.  Radial velocity follow-up studies have been performed and orbital periods are being determined in $\sim$ 25 systems (N. Silvestri 2006, priv. comm.).  \citet{wac03} have searched the 2MASS point source catalog \citep{cut03} for WDs with infrared excess, which may indicate the presence of an unresolved red dwarf companion, and identified 95 systems.  They have followed up these identifications with Hubble Space Telescope (HST) snapshot observations to determine if indeed these WDs have low mass companions.  First results from these HST observations \citep{far06} indicate that of the 48 systems studied thus far, 27 have fully or partially resolved companions and 15 more are almost certainly unresolved binaries.  The resolved systems have estimated projected orbital separations between $\sim$ 10 - 1000 AU, making it unlikely that any of these systems underwent a CE phase.  Estimated upper limits for the unresolved binaries suggest orbital separations less than $\sim$ 5 AU.   \citet{far06} suggest that these unresolved systems are likely WD + red dwarf pairs in close orbits and may be good candidates for radial velocity variables.  Assuming the 48 systems studied thus far are representative of the entire sample of 95, they expect to resolve $\sim$ 55 wide systems and identify a total of $\sim$ 30 unresolved close binaries for future radial velocity studies.

\subsubsection{Compilations} 

\citet{sch03} compiled a sample of 30 well-studied PCEBs with known orbital periods.  Since they were primarily interested in PCEBs that would become CVs within a Hubble time, they restricted their sample to systems that have mass ratios less than one (i.e., $M_s$ $\le$ $M_{WD}$), have orbital periods shorter than 2 days, and have main sequence (or BD) secondaries (i.e., no subgiant or giant secondary stars).  More recently, \citet{mor05} compiled a list of all detached binaries containing at least one WD and having known orbital periods. 
In their compilation, 32 binaries are confirmed WD + red dwarf pairs and 9 are sdOB + red dwarf pairs.  Their sample includes 28 of the systems in the \citet{sch03} compilation.  To the best of our knowledge, the \citet{mor05} compilation provides the most extensive list of PCEBs with known orbital periods that is currently available, although once orbital periods are determined for the subset of systems from SDSS and 2MASS, that number should increase significantly.  

Regrettably, neither the PCEBs in the \citet{mor05} compilation nor those culled from the large SDSS and 2MASS data releases provide a well-defined, statistically complete sample of PCEBs.  For example, the majority of the systems in the \citet{sil06} SDSS sample were targeted for spectroscopy because their colors resembled other, higher priority objects such as quasars (see \citealp{ric02} for additional details).  In addition, PCEBs containing either a hot WD and a late M/L type secondary or a cool WD and an early M/late K type secondary present particular observational challenges and have likely been missed generally (see \citealp{sch03} for a nice discussion of this).  In the former case, the WD is much brighter than the secondary in the optical and the system may be mistaken for a single WD based on its colors.  In the latter case, the WD is hidden by the glare of the secondary and the system may be misidentified as a single early M or K dwarf based on its colors.  A proper sample of PCEBs for comparison with population models such as those presented in this paper does not exist as yet.  Until it does, any comparison with theoretical populations of PCEBs would be premature.  However, the rapid growth in the number of observed PCEBs over just the past few years suggests that such a sample may be feasible within the next several years.  Efforts are already in progress to improve the observed sample of PCEBs over the next 2-3 years (B. G\"ansicke, priv. comm. 2006).

\subsection {Observations of CVs}

In the latest online version (RKcat7.6, 1 January 2006) of the \citet{rit03} catalog, there are 604 CVs with known orbital periods (we exclude systems that RK designate as possible AM CVn stars).   The distribution of orbital periods in observed CVs provides, by far, the best test of CV population models.  We show this distribution in Figure 10.  It is important to remember that the CVs in Figure 10 were born throughout the Galaxy's history and therefore do not represent present-day ZACVs.  Consequently, because of secular evolution, the distribution of orbital periods shown in Fig. 10 is most likely different than the distribution of orbital periods in ZACVs (see e.g., \citealp{kol93,how01}).   Nevertheless, there are some features of the observed orbital period distribution in CVs that may provide clues as to the ZACV orbital period distribution and, therefore, are germane to the present discussion.

\subsubsection {The scarcity of CVs with orbital periods shorter than 77 min:  evidence for a cutoff mass?}

One of the most prominent features in the observed orbital period distribution of CVs is the existence of a minimum period at 77 min.  Of the $\sim$ 600 CVs in this distribution, only 3 have orbital periods shorter than 77 min:  V485 Cen (59 min), EI Psc (64 min), and J1507+5230 (67 min).  According to the accepted theoretical explanation of the period minimum \citep{pac81,rap82}, CVs evolve to shorter orbital periods until at some point the secondary has lost so much mass that its thermal timescale becomes longer than the mass transfer timescale.  This point roughly coincides with the secondary having lost so much mass that it has also become substellar.  Once this point is reached, the secondary's response to mass loss changes and it now expands upon mass loss rather than contracts.  Further mass transfer then results in an increase in the orbital period of the system \citep{pac81,rap82}.  Consequently, CVs that form with orbital periods longer than the orbital period minimum cannot evolve to orbital periods shorter than it.  

This explanation of the period minimum is suitable for CVs that form with orbital periods greater than the period minimum, but it does not address CVs that form with orbital periods shorter than the period minimum.  \citet{pol04} performed the first detailed population synthesis calculations of the formation of CVs with ultrashort orbital periods.\footnote{We note that AM CVn stars have ultrashort orbital periods, but these systems have helium-rich donors.  By ultrashort-period CVs, we are specifically referring to systems containing hydrogen-rich donors that have orbital periods below 77 min.}.  As noted in \citet{pol04}, the secondaries in CVs that form with orbital periods less than 77 min are very low mass or substellar.  Consequently, realistic population synthesis calculations of the formation of ultrashort-period CVs awaited the development of reliable detailed models of very low mass stars and BDs over the past five to ten years (e.g., \citealp{cb00}). 

Our standard model uses essentially the same input physics for very low mass/substellar stars as in \citet{pol04} (see section 2), and it predicts that almost 1/4 of all present-day ZACVs form with orbital periods shorter than 77 min (see Table 3 and Figure 6).  Where are these systems?  Or to rephrase the question, how do we explain the current lack of CVs with orbital periods shorter than 77 min?  We suggest four possible answers: (1) all CVs ever formed below 77 min have secularly evolved to orbital periods longer than 77 min by the present epoch; (2) CVs formed below the period minimum are too faint to see in quiescence and have recurrence lifetimes too long to have seen an outburst in recent history; (3) the assumption in our standard model that $\alpha_{CE}$ = 1 is not valid; and (4) the assumption in our standard model that $g(q)$ is flat is not valid.  

The first possibility is unlikely.  \citet{kol99} calculated secular evolution sequences for CVs that formed with a 0.6 $\Msun$ WD and secondary masses between 0.04 and 0.21 $\Msun$ using full stellar and substellar models for the secondary.  These CVs formed at orbital periods between 54 min and 2.1 hr.  For a 0.07 $\Msun$ secondary, they found that it took 1.5 Gyr for the system to evolve from 54 min to 77 min.\footnote{In the \citet{kol99} study, gravitational radiation was the only angular momentum loss mechanism driving secular evolution. If an additional source of angular momentum loss is allowed to operate below the gap (e.g., \citealp{pat98,wil05}), this timescale would be reduced.}  Given that almost 1/4 of all CVs are predicted to form with orbital periods less than 77 min at the present epoch, it is difficult to understand how subsequent secular evolution could evacuate the entire orbital period regime shorter than 77 min.  As a comparative example, a CV that formed at the upper edge of the period gap would take $\sim$ 1 Gyr to cross the gap, assuming only gravitational radiation (e.g., \citealp{how97}).  Yet the period gap is not completely devoid of CVs.

The second possibility is also unlikely.  \citet{kol99} also computed mass transfer rates as a function of orbital period for the same evolutionary sequences discussed in the previous paragraph.  They found that the evolutionary tracks merge into a common track at an orbital period of $\sim$ 70 minutes regardless of whether the CV was formed with an orbital period longer than 77 min or shorter than 77 min (see Fig. 1 in \citealp{kol99}).  Indeed, their models predict that present-day CVs with orbital periods less than 77 min should have mass transfer rates greater than or equal to present-day CVs with orbital periods slightly greater than or equal to 77 min.  Well known examples of the latter are WZ Sge and AL Com (P = 81 min for both systems).  These two systems and other WZ Sge-type systems were not discovered in quiescence, but were discovered because of their outbursts.  Consequently, if dwarf nova outbursts have been observed in these systems, then it is difficult to understand why similar outbursts would not have been observed in systems just below the period minimum, assuming they exist.  Since such systems are predicted to have mass transfer rates similar to or greater than WZ Sge-type systems \citep{kol99}, they should have recurrence times that are similar or even shorter.   

Determining the viability of the third possibility is one of our goals in undertaking the present investigation.  To that end, inspection of Table 3 shows that the fraction of ZACVs with orbital periods shorter than 77 min is zero in 2 of the 12 models:  PL 2.0 and CUTM 0.15.  A common feature of these models is that $\alpha_{CE}$ is exactly or essentially zero below some secondary mass. For the CUTM 0.15 model, that mass is 0.15 $\Msun$.  For the PL 2.0 model, that mass is $\sim$ 0.10 $\Msun$.  Consequently, these two models demonstrate that if merger within the CE is inescapable below a secondary mass of $\sim$ 0.1 $\Msun$ or greater, agreement between the predicted and the observed orbital period distribution in CVs below 77 min can be greatly improved.  We note that in 3 of the models, CON 0.05, PL 1.0 and CUTM 0.075, the fraction of CVs formed below 77 minutes is 0.05.  While this is still an order of magnitude higher than the observed fraction (at best, 3 out 604 or $\sim$ 0.005), these may be viable models if subsequent secular evolution can move systems to periods longer than 77 min on a sufficiently short time scale.

The fourth possibility is that the distribution of ZAMS binaries containing a MS primary and a very low mass/BD secondary is different than the distribution of ZAMS binaries containing two MS stars.  Observations of companions to solar-type stars (spectral types F-M) indicate a relative scarcity of BD companions compared with either less massive planetary companions \citep{mar00} or more massive stellar companions (e.g., \citealp{duq91,fis92}). This comparative deficit of BD companions to solar-type stars has been termed the ``brown dwarf desert.''  Recently, \citet{gre06} have described quantitatively the characteristics of the brown dwarf desert for nearby binaries ($d <$ 50 pc) with orbital periods less than 5 yr (orbital separations $\lesssim$ 3 AU).  They find that $\sim$ 16\% of these nearby binaries contain solar type-stars with companions more massive than Jupiter:  $\sim$ 11 \% have stellar companions, $<$ 1\% have BD companions, and $\sim$ 5\% have planetary companions.  \citet{pol04} found that  all ZACVs with BD secondaries evolved from progenitor ZAMS binaries with orbital separations less than 3 AU and $\sim$ 75\% of them evolved from progenitor ZAMS binaries with F-type or later primaries.  This places the majority of the ZAMS binary progenitors of ZACVs with BD secondaries within the brown dwarf desert.  Therefore, it is also possible that the lack of observed CVs with orbital periods shorter than the period minimum may be due to a lack of progenitor binaries from which to draw.  Interestingly, if hydrodynamical models of the CE phase eventually do not indicate the existence of a cutoff mass $\ga$ 0.1 $\Msun$, this could provide strong, independent support for the existence of the brown dwarf desert.

\subsubsection{The lower edge of the period gap: an imprint of CV formation?}

\citet{wil05} have computed comprehensive model populations of present-day CVs that formed with orbital periods less than 2.75 hr, including models with an additional source of angular momentum loss besides gravitational radiation.  As an example of such a source, they considered circumbinary disks.   Their model populations that include circumbinary disks successfully resolve the well-known discrepancy between the predicted and the observed value of the period minimum in CVs (see \citealp{pat98} and \citealp{kol99} for a discussion of this discrepancy).  However, these models fail to reproduce the sharp lower edge of the period gap at 2.25 hr seen in the observed distribution.  Willems et al. note that agreement could be improved by ``a significant modification of the ZACV population near the lower edge of the period gap'' or by a flow of systems into the gap from CVs formed above the gap.   They simulate the latter flow by artifically multiplying the birthrate at 2.25 hr in their models by a factor of $\sim$ 100.  However, as shown in a very recent follow-up paper \citep{wil07}, this flow is inconsistent with circumbinary disk models of CVs that form above the gap.  \citet{wil07} find that the additional angular momentum loss created by the circumbinary disk increases the mean mass transfer rate in CVs formed above the gap and causes these systems to bounce at an orbital period of 2.75 hr (the upper edge of the gap).  Therefore, CVs that formed above the gap in their models with circumbinary disks will always remain above the gap and, thus, cannot create the flow of systems to orbital periods shorter than 2.75 hr required by \citet{wil05}.  Consequently, it would seem that better agreement between their models and observations concerning the lower edge of the period gap must be achieved by a modification of the standard model for the ZACV population. 

In our PL model sequence for ZACVs, we find that as the value of $n$ is increased, the fraction of CVs formed below the period gap decreases and the fraction of CVs formed in the gap increases (see Table 3).  This trend is especially evident in the PL 2.0 model, where the fraction of CVs that form below the gap is reduced by over a factor of 7 and the fraction of CVs that form in the gap is increased by over a factor of 4 compared with our standard model.  In addition, we find that as $n$ is increased, the short-period peak in the orbital period distribution of ZACVs shifts to longer orbital periods (see Fig. 8).  For our standard model, this peak occurs at an orbital period of 1.8 hr (log P = -1.125 in Fig. 6).  For the PL 0.5, 1.0 and 2.0 models, this peak occurs at an orbital period of 1.9 hr (log P = -1.10), 2.1 hr (log P = -1.05), and 2.4 hr (log P = -1.0), respectively.  Interestingly then, this shift provides precisely the modification of the ZACV population required by \citet{wil05}.  We note that in the CUTM model sequence, increasing the cutoff mass also shifts the short-period peak in the orbital period distribution to longer orbital periods, although to a lesser degree.  For the CUTM 0.15 model, this peak occurs at 2.0 hr (log P = -1.175 in Fig. 9).  Extrapolating from our CUTM model sequence, for this peak to occur at 2.25 hr, a cutoff mass of 0.27 $\Msun$ would be required.

We suggest, therefore, that it is possible that the sharp lower edge of the period gap may not be due solely to secular evolution, but could also be, in part, an imprint of CV formation.  In the standard model of CV secular evolution \citep{kin88}, systems formed above the gap evolve to shorter orbital periods on a rapid timescale due to magnetic braking.   Such rapid evolution effectively washes out any features from the underlying ZACV orbital period distribution (see \citealp{kol93,how01}), making such an imprint unlikely.  However, recent observations of magnetic activity in single stars in young open clusters \citep{sil00, pin02} and of X-ray emission from rotating dwarfs \citep{piz03} have called into question the magnetic braking prescription used in the standard model (disrupted magnetic braking, e.g., \citealp{rap83}).  Magnetic braking prescriptions developed by \citet{and03} based on the open cluster observations and by \citet{iva03} based in part on the X-ray emission data produce angular momentum loss rates above the gap that are significantly lower than in standard magnetic braking.  It is possible, therefore, that in a model of secular evolution where the rapid flow of systems from above to below the gap is reduced (as in the prescriptions by Andronov et al. and Ivanova \& Taam) or eliminated (as in secular evolution with circumbinary disks), features in the ZACV orbital period distribution may persist and be evident in the present-day (evolved) CV distribution.\footnote{We note that \citet{web79} made a somewhat similar argument in his so-called ``static'' explanation of the period gap.}  In order for the lower edge of the period gap to be such an imprint, the standard model of CV formation must be modified to produce a peak in the ZACV orbital period distribution at $\sim$ 2.25 hr.  One way of making such a modification is to assume that $\alpha_{CE}$ is a moderate to strong function of secondary mass (e.g., $n \ga$ 1 in PL model sequence).  However, we note that our preliminary investigation of the dependence of $\alpha_{CE}$ on the evolutionary state of the giant indicates that such a peak can be achieved in other ways as well.

\subsection {The dependence of $\alpha_{CE}$ on other orbital parameters}

It is very unlikely that $\alpha_{CE}$ depends solely on secondary mass.  As a simple example, if the efficiency of energy transfer depends upon the magnitude of the drag torque, then $\alpha_{CE}$ would not only depend upon the secondary mass, but also upon the density profile within the giant.  This profile is determined by the core mass and total mass of the giant.  Thus, $\alpha_{CE}$ would be a multi-dimensional function of $M_s$, $M_c$, and $M_p$ in this scenario.  In general, we would expect that $\alpha_{CE}$ is at least a function of secondary mass and the evolutionary state of the giant star, which is primarily determined by the giant's core mass.  For example, $\alpha_{CE}$ might be different for CE evolution involving a first giant branch primary than for CE evolution involving an AGB primary (see e.g., \citealp{san00, wil05}).   To address the potential multi-dimensional nature of $\alpha_{CE}$, we have chosen to investigate the dependence of $\alpha_{CE}$ upon secondary mass in this first paper and its dependence upon the evolutionary state of the giant in paper II.  Clearly, we can only hope to approximate the true dependence of $\alpha_{CE}$ on the orbital parameters of the binary by considering its dependence on each parameter separately.  Therefore, we urge caution when interpreting our results, particularly when it comes to inferring a particular $\alpha_{CE}(M_s)$ dependence based on comparison with observations of PCEBs and ZACVs.  Observed trends in a given orbital parameter distribution for either population could also be due to strong dependences of $\alpha_{CE}$ on orbital parameters other than secondary mass, for example.

\subsection {Guiding detailed theoretical models of the CE phase and future observations of PCEBs}

One of our main goals in conducting the present study is to identify interesting regions of orbital parameter space that deserve further study by researchers who are calculating detailed hydrodynamical models of the CE phase.   As mentioned in section 1, such calculations are extremely time intensive and demand an efficient use of available computational resources.  To that end, we believe that calculating hydrodynamical models for a fixed, typical giant mass and a sequence of low mass secondaries should be given some priority.  Our model populations of ZACVs indicate that if $\alpha_{CE}$ is very low ($\lesssim$ 0.20), the distribution of ZACVs with orbital periods less than $\sim$ 3 hr will be significantly altered compared with our standard model in which $\alpha_{CE}$ = 1.0 (see Fig. 6).  Such a modification of the ZACV population may have important ramifications for our understanding of the orbital period distribution in CVs below and in the period gap.  As mentioned in section 1, $\alpha_{CE}$ may be very small for CE evolution involving a very low mass secondary.  In such cases, the rate of orbital energy dissipation within the CE is becoming comparable to the luminosity of the giant and insufficient to drive the bulk hydrodynamical motion needed the eject the envelope (R. Taam, 2003, private communication).  However, there have only been two detailed hydrodynamical calculations of CE evolution involving a secondary with $M_s \leq$ 0.30 $\Msun$ and both involved giants with core masses $\leq$ 0.45 $\Msun$ (i.e. on the first giant branch) \citep{san00}.  

We suggest calculating a sequence of models with an AGB primary that has a 0.6 $\Msun$ core and secondaries ranging in mass from 0.05 $\Msun$ to 0.30 $\Msun$.  The number of models calculated will depend upon the available computational resources, but should include a sufficient number of models to estimate a dependence of $\alpha_{CE}$ on secondary mass.  At a minimum, models with $M_s$ = 0.05, 0.10, 0.20 and 0.30 $\Msun$ should be calculated.   If the evolution involving the 0.10, 0.20 and 0.30 differ significantly, then additional models with 0.15 $\Msun$ or 0.25 $\Msun$ may need to be calculated.  

If the above sequence of models indicates the existence of a cutoff mass for the secondary below which merger always occurs, then we suggest that additional CE models with that cutoff mass be calculated for other giant masses.  If the value of the cutoff mass is found not to be strongly dependent upon the mass of the giant star, then such a cutoff mass may provide a prediction of hydrodynamical models of CE evolution that is feasible to test observationally.  The chances of observing a binary in the CE phase are miniscule since the phase is so short-lived, making direct tests of detailed model calculations impossible.  Consequently, we must rely on observations of PCEBs to provide indirect tests of hydrodynamical models of the CE phase.  The value of a cutoff mass predicted from these hydrodynamical models does not require population synthesis calculations of present-day PCEBs in order to test it.  Such population synthesis calculations are inherently dependent upon uncertain quantities such as the prescription for angular momentum loss following the CE phase (i.e., magnetic braking) and the mass ratio distribution in ZAMS binaries.  Further, such calculations require a prescription for the CE phase, not just the value of an extremum for one orbital parameter.   For these reasons, a predicted value of a cutoff secondary mass, assuming one exists, is clearly advantageous and would provide a direct, basic test of hydrodynamical models of CE evolution.

Observationally, providing a sample of PCEBs that is both well-defined and statistically complete, especially with regard to very low mass secondaries, should be given some priority.  Regardless of whether or not a cutoff mass for the secondary exists, such a sample is crucially needed if we are to have any confidence in the results of hydrodynamical calculations of CE evolution.

\section{Conclusions}

We have undertaken a systematic study of the effects of a CE efficiency parameter that is a function of secondary mass on models of the present-day populations of PCEBs and ZACVs and their distributions over orbital period, orbital separation, secondary mass and WD mass.  We investigated three possibilities: (1) constant $\alpha_{CE}$ models for six values of $\alpha_{CE}$:  1.0, 0.6, 0.3, 0.2, 0.1, and 0.05, with $\alpha_{CE}$ = 1 designated as our standard model, (2) a power law dependence, $\alpha_{CE}$ = $(M_s)^n$, with $n$ = 0.5, 1.0 and 2.0, and (3) a dependence in which $\alpha_{CE}$ approaches 1 for large secondary masses and $\alpha_{CE}$ = 0 below some assumed cutoff mass, $\alpha_{CE} = 1-M_{cut}/M_s$, with $M_{cut}$ = 0.0375, 0.075 and 0.15 $\Msun$.  We denoted these constant, power law and cutoff mass model sequences by CON, PL, and CUTM, respectively.

The two most important findings of this study are: 

(1)  Our CON model sequence for ZACVs shows that for values of $\alpha_{CE}$ $\lesssim$ 0.2, the predicted population of present-day ZACVs differs significantly from our standard model.  All prior population synthesis studies of the formation of CVs only considered values of $\alpha_{CE}$ $\geq$ 0.3 \citep{pol88,pol96,dek92,how01,wil05} and found that their model populations were not strongly dependent upon the value of $\alpha_{CE}$.    Our results indicate that a much wider range of values for $\alpha_{CE}$, including very low values, must be considered in order for a dependence to be seen.  

(2) Our model sequences for ZACVs in which $\alpha_{CE}$ is a function of secondary mass show that if $\alpha_{CE}$ decreases rapidly for small secondary masses, the orbital period distribution in ZACVs for orbital periods in and below the gap differs significantly from our standard model.  These differences are most evident in the PL 2.0 and CUTM 0.15 models.  For these models, the fraction of CVs forming with orbital periods below the gap is reduced significantly, the fraction forming in the gap is increased significantly, and both the short-period peak and the minimum period in the ZACV orbital period distribution shift to considerably longer orbital periods.  These results indicate that if $\alpha_{CE}$ is strongly dependent upon $M_s$ for small secondary masses, the standard model for the formation of CVs \citep{dek92, pol96} with orbital periods below $\sim$ 3 hr will need to be modified.

Other findings of this study are:

(3)  Approximately 25\% of all present-day ZACVs in our standard model are predicted to form with orbital periods shorter than 77 min, the observed orbital period minimum in CVs.  These ultrashort-periods occur only in CVs that form with very low mass (late M) or substellar secondaries.  Even with subsequent secular evolution, this prediction is in sharp conflict with the observed percentage of CVs with $P <$ 77 min, which is $<$ 1\%.  In our PL 2.0 and CUTM 0.15 models, no CVs are formed with $M_s \lesssim$ 0.10 $\Msun$ because of the steep decline in $\alpha_{CE}$ for very low secondary masses.  Consequently, no CVs are formed with $P <$ 77 min in these models, in much better agreement with the observations.  This suggests the possibility that the observed scarcity of CVs with $P <$ 77 min may be evidence that a cutoff secondary mass for CE evolution exists and is $\ga$ 0.10 $\Msun$.  However, at the moment this is not clear, since this scarcity of ultrashort-period CVs could also be due to other factors such as a lack of CV progenitors with BD secondaries.  If hydrodynamical models of the CE phase eventually do not indicate the existence of a cutoff mass $\ga$ 0.1 $\Msun$, this could provide strong, independent support for the existence of the brown dwarf desert.

(4) It is possible that the sharp lower edge of the period gap at 2.25 hr could be, in part, an imprint of CV formation.  However, two key modifications of the standard model of CV formation and evolution are required in order for this to be true.  First, the rapid secular evolution of systems to shorter orbital periods from above the gap must be significantly reduced or eliminated.  This would require a modification of standard magnetic braking similar to those proposed recently by \citet{and03}, \citet{iva03}, and \citet{wil05,wil07}.  Second, the short-period peak in the ZACV orbital period distribution must be shifted from $\sim$ 1.80 hr to $\sim$ 2.25 hr.  Such a shift can be produced if $\alpha_{CE}$ is moderately to strongly dependent upon $M_s$ ($n \ga$ 1 in our PL model sequence).  Preliminary investigation of the dependence of $\alpha_{CE}$ on the state of the giant star, which we will report upon in paper II, indicates that this shift can produced in other ways as well. 

(5) The fraction of systems that survive the CE phase, $f_{surv}$, in our CON model sequence decreases approximately as $(\alpha_{CE})^{0.5}$ from 0.50 to 0.07 as $\alpha_{CE}$ is decreased from 1.0 to 0.05 (see Table 1).  In our PL and CUTM model sequences, $f_{surv}$ remains $\lesssim$ 1/2 for all models and varies rather modestly from 0.39 to 0.48. 

(6) Over a wide range of values for $\alpha_{CE}$ in our CON sequence, our models predict that the majority of present-day PCEBs contain secondaries with masses $>$ 0.375 $\Msun$, orbital periods $>$ 1 day and orbital separations $>$ 0.025 AU, with most having periods of $\sim$ 3 days and separations of $\sim$ 0.05 AU.  These models further predict that the present-day population of PCEBs should contain roughly an equal number of systems with He and CO WDs if $\alpha_{CE}$ is globally near unity, but should be clearly dominated by systems containing CO WDs if $\alpha_{CE}$ is globally small ($\lesssim$ 0.30).

(7) In our PL and CUTM model sequences for present-day PCEBs, the only distribution that varies significantly is the secondary mass distribution.  In the PL sequence, as $n$ is increased from 0.5 to 2.0, the fraction of PCEBs with low-mass secondaries decreases by a factor of $\sim$ 4 from 0.26 to 0.06.  The decrease in the number of low-mass secondaries is particularly severe for the $n$ = 2.0 model.  In this case, there are no present-day PCEBs with secondary masses less than $\sim$ 0.10 $\Msun$.  In the CUTM model sequence, significant changes only occur near the cutoff mass and the distributions are nearly identical for $M_s \ga$ 0.375 $\Msun$. 

(8) In our CON model sequence for present-day ZACVs, if $\alpha_{CE}$ $\ga$ 0.30, the majority of the systems contain He WDs and have orbital periods below the period gap.  On the other hand, if $\alpha_{CE}$ $\lesssim$ 0.20, then the majority of the systems contain CO WDs and there is a significant shift in the orbital period distribution towards periods in and above the gap. The majority of ZACVs contain low-mass secondaries regardless of the value of $\alpha_{CE}$.

(9) In our PL model sequence of present-day ZACVs, the majority of the systems contain He WDs regardless of whether the dependence of $\alpha_{CE}$ on $M_s$ is weak ($n$ = 0.5) or strong ($n$ = 2.0).  Systems with low-mass secondaries dominate if the dependence of $\alpha_{CE}$ on $M_s$ is weak to moderate ($n \leq$ 1), whereas systems with high-mass secondaries dominate if the dependence is strong.  Finally, systems with orbital periods below the period gap are favored if the dependence of $\alpha_{CE}$ on $M_s$ is weak, while systems in and above the period gap are favored if the dependence is moderate to strong. 

(10) In our CUTM model sequence of present-day ZACVs, the majority of the systems contain low-mass secondaries and have orbital periods below the period gap if $M_{cut} \leq$ 0.075 $\Msun$, while the majority of the systems contain high-mass secondaries and have orbital periods above the gap if $M_{cut}$ = 0.15 $\Msun$.  Further, approximately the same number of systems contain He WDs and CO WDs if $M_{cut} \leq$ 0.075 $\Msun$, while the majority of the systems contain CO WDs if $M_{cut}$ = 0.15 $\Msun$.

(11) One of our main goals in conducting the present investigation is to provide guidance to those performing detailed hydrodynamical calculations of CE evolution.  To that end, we suggest that calculating detailed models for a fixed mass giant and a sequence of very low mass secondaries, ranging from 0.3 $\Msun$ down to 0.05 $\Msun$, should be given some priority.  The results of our study indicate that understanding CE evolution involving very low mass secondaries may have important ramifications for our understanding of the formation of CVs in and below the period gap.  In addition, if hydrodynamical models of the CE phase determine that a cutoff secondary mass exists and is not strongly dependent upon the core mass of the giant star, then the value of such a cutoff mass could provide a prediction of these models that is independent of any population synthesis calculations and is feasible to test observationally.  

(12) We further suggest that observing a well-defined sample of PCEBs that is statistically complete, particularly with regard to very low mass secondaries, should be given some priority.  Such a sample will provide important observational tests of hydrodynamical models of the CE phase.

\acknowledgements {Warm thanks to C. Carden, O. De Marco, D. Drapes, B. G\"{a}nsicke, D. Hoard, K. Simkunas, R. Taam, and B. Willems for very useful discussions about this work.  We also thank the anonymous referee for comments that improved the quality and the clarity of the paper.  This work was funded in part by NSF grant AST-0328484 and by a grant from the Wisconsin Space Grant Consortium to Marquette University.}

\clearpage

\begin{deluxetable}{lc}
\tabletypesize{\scriptsize}
\tablecaption{Fraction of systems, $f_{surv}$, that survive CE evolution}
\tablewidth{0pt}
\tablehead{
\colhead{Model} & \colhead{$f_{surv}$}
}
\startdata
CON 1.0  & 0.50 \\
CON 0.6  & 0.41 \\
CON 0.3  & 0.28 \\
CON 0.2  & 0.22 \\
CON 0.1  & 0.13 \\
CON 0.05 & 0.07 \\
PL 0.5   & 0.46 \\
PL 1.0   & 0.43 \\
PL 2.0   & 0.39 \\
CUTM 0.0375 & 0.48 \\
CUTM 0.075  & 0.46 \\
CUTM 0.15   & 0.43 \\
\enddata
\end{deluxetable}

\clearpage

\begin{deluxetable}{lcccccc}
\tablecolumns{7}
\tabletypesize{\scriptsize}
\tablecaption{Selected results for model populations of PCEBs}
\tablewidth{0pt}
\tablehead{
\colhead{Model} & \multicolumn{4}{c}{Fraction with} & \colhead{} & \colhead{$N/N_{stand}$} \\
\cline{2-5} \\
\colhead{} & \colhead{He WDs} & \colhead{CO WDs} & 
\colhead{$M_s <$ 0.375 $\Msun$} & \colhead{$P <$ 1 day} & 
}
\startdata
CON 1.0 & 0.51 & 0.49 & 0.36 & 0.23 & & 1.00 \\
CON 0.6 & 0.43 & 0.57 & 0.35 & 0.25 & & 0.74 \\
CON 0.3 & 0.31 & 0.69 & 0.33 & 0.27 & & 0.45 \\
CON 0.2 & 0.23 & 0.77 & 0.31 & 0.27 & & 0.32 \\
CON 0.1 & 0.10 & 0.90 & 0.26 & 0.24 & & 0.17 \\
CON 0.05 & 0.02 & 0.98 & 0.20 & 0.19 & & 0.09 \\
PL 0.5 & 0.45 & 0.55 & 0.26 & 0.19 & & 0.80 \\
PL 1.0 & 0.40 & 0.60 & 0.18 & 0.15 & & 0.66 \\
PL 2.0 & 0.34 & 0.66 & 0.06 & 0.08 & & 0.50 \\
CUTM 0.0375 & 0.51 & 0.49 & 0.33 & 0.22 & & 0.92 \\
CUTM 0.075 & 0.47 & 0.53 & 0.29 & 0.20 & & 0.85 \\
CUTM 0.15 & 0.43 & 0.57 & 0.21 & 0.15 & & 0.72 \\
\enddata
\end{deluxetable}

\clearpage

\begin{deluxetable}{lcccccccc}
\tabletypesize{\scriptsize}
\rotate
\tablecaption{Selected results for model populations of ZACVs}
\tablewidth{0pt}
\tablehead{
\colhead{Model} & \multicolumn{7}{c}{Fraction with} & \colhead{$N/N_{stand}$} \\ 
\cline{2-8} \\
\colhead{} & \colhead{He WDs} & \colhead{CO WDs} & 
\colhead{$M_s <$ 0.375 $\Msun$} & \colhead{$P <$ 77 min} & \colhead{$P <$ 2.25 hr} & \colhead{2.25 hr $< P <$ 2.75 hr} & \colhead{$P >$ 2.75 hr} &  
}
\startdata
CON 1.0 & 0.58 & 0.42 & 0.72 & 0.24 & 0.67 & 0.04 & 0.29 & 1.00 \\
CON 0.6 & 0.54 & 0.46 & 0.72 & 0.23 & 0.65 & 0.05 & 0.29 & 0.96 \\
CON 0.3 & 0.46 & 0.54 & 0.69 & 0.17 & 0.58 & 0.09 & 0.34 & 0.83 \\
CON 0.2 & 0.41 & 0.59 & 0.68 & 0.15 & 0.53 & 0.11 & 0.35 & 0.72 \\
CON 0.1 & 0.26 & 0.74 & 0.61 & 0.10 & 0.41 & 0.15 & 0.44 & 0.48 \\
CON 0.05 & 0.12 & 0.88 & 0.51 & 0.05 & 0.27 & 0.16 & 0.56 & 0.28 \\
PL 0.5 & 0.46 & 0.54 & 0.67 & 0.15 & 0.58 & 0.07 & 0.35 & 0.84 \\
PL 1.0 & 0.31 & 0.69 & 0.57 & 0.05 & 0.42 & 0.12 & 0.46 & 0.63 \\
PL 2.0 & 0.08 & 0.92 & 0.35 & 0.00 & 0.09 & 0.18 & 0.72 & 0.37 \\
CUTM 0.0375 & 0.53 & 0.47 & 0.68 & 0.18 & 0.62 & 0.05 & 0.34 & 0.86 \\
CUTM 0.075 & 0.46 & 0.54 & 0.63 & 0.05 & 0.54 & 0.06 & 0.39 & 0.74 \\
CUTM 0.15 & 0.28 & 0.72 & 0.48 & 0.00 & 0.31 & 0.13 & 0.55 & 0.52 \\
\enddata
\end{deluxetable}

\clearpage


\begin{figure}
\plotone{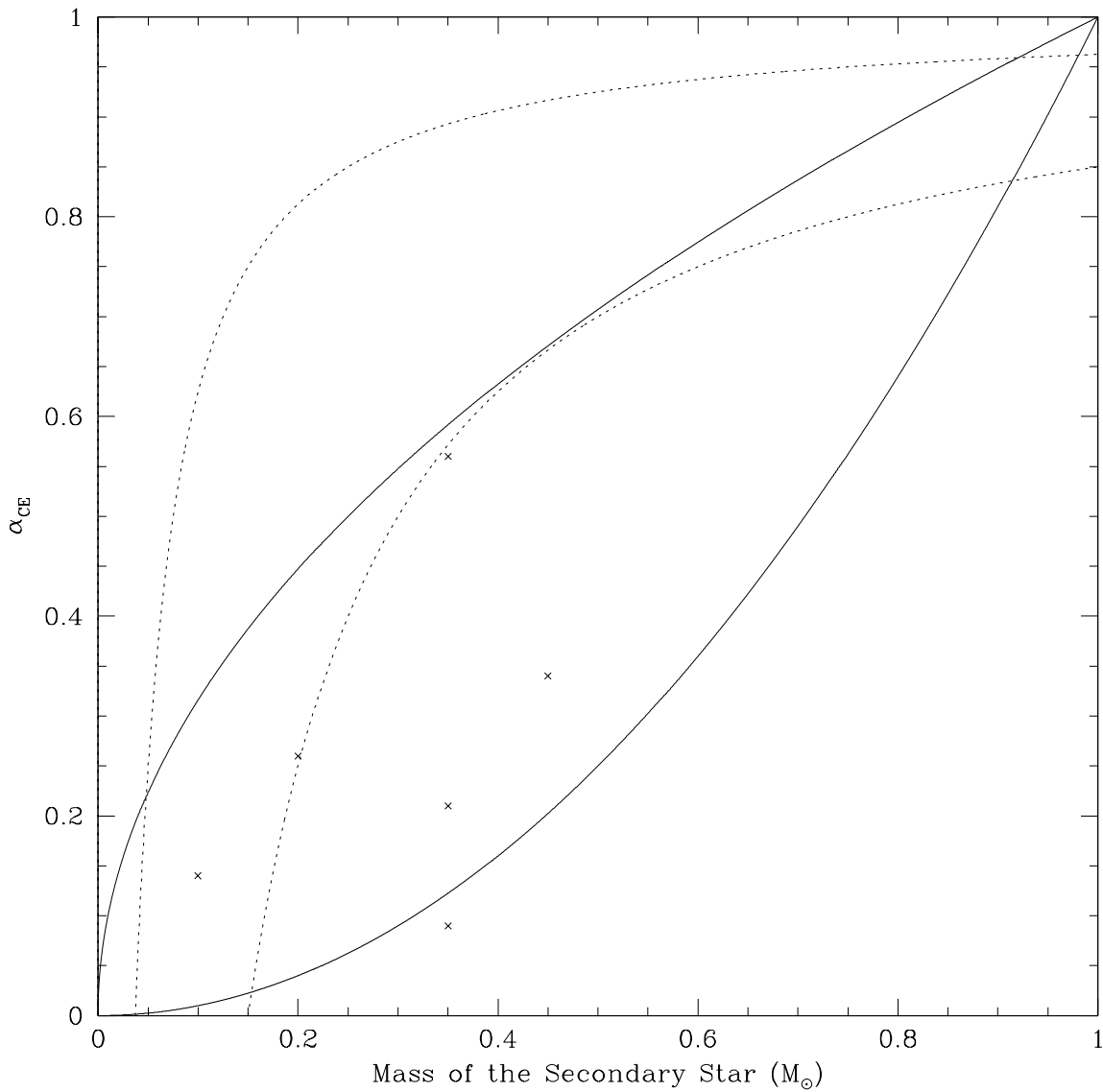}
\figcaption{Comparison of assumed functional forms for $\alpha_{CE}$ with estimated values of $\alpha_{CE}$ from detailed hydrodynamical calculations (crosses).  The solid lines show $\alpha_{CE}$ = $(M_s)^n$ for $n$ = 0.5 (top curve) and $n$ = 2.0 (bottom curve).  The dotted lines show $\alpha_{CE} = 1-M_{cut}/M_s$ for $M_{cut}$ = 0.0375$\Msun$ (top curve) and $M_{cut}$ = 0.15$\Msun$ (bottom curve).}
\end{figure}

\begin{figure}
\plotone{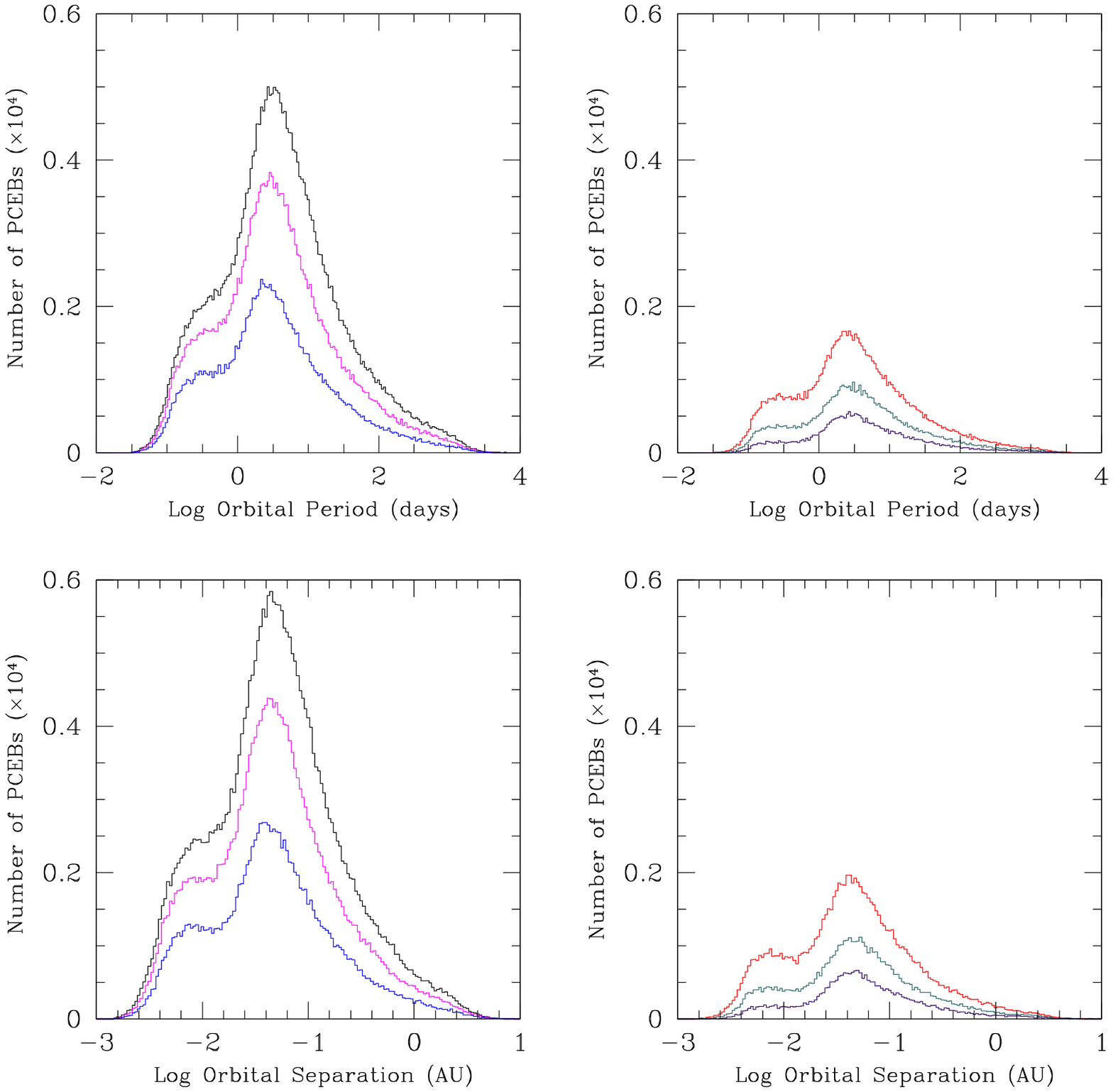}
\figcaption{Theoretical present-day distributions of the orbital periods (top panels) and the orbital separations (bottom panels) in PCEBs for $\alpha_{CE}$ = 1.0 (black), 0.6 (magenta), and 0.3 (blue) (left panels) and for $\alpha_{CE}$ = 0.2 (red), 0.1 (green) and 0.05 (purple) (right panels).}
\end{figure}

\begin{figure}
\plotone{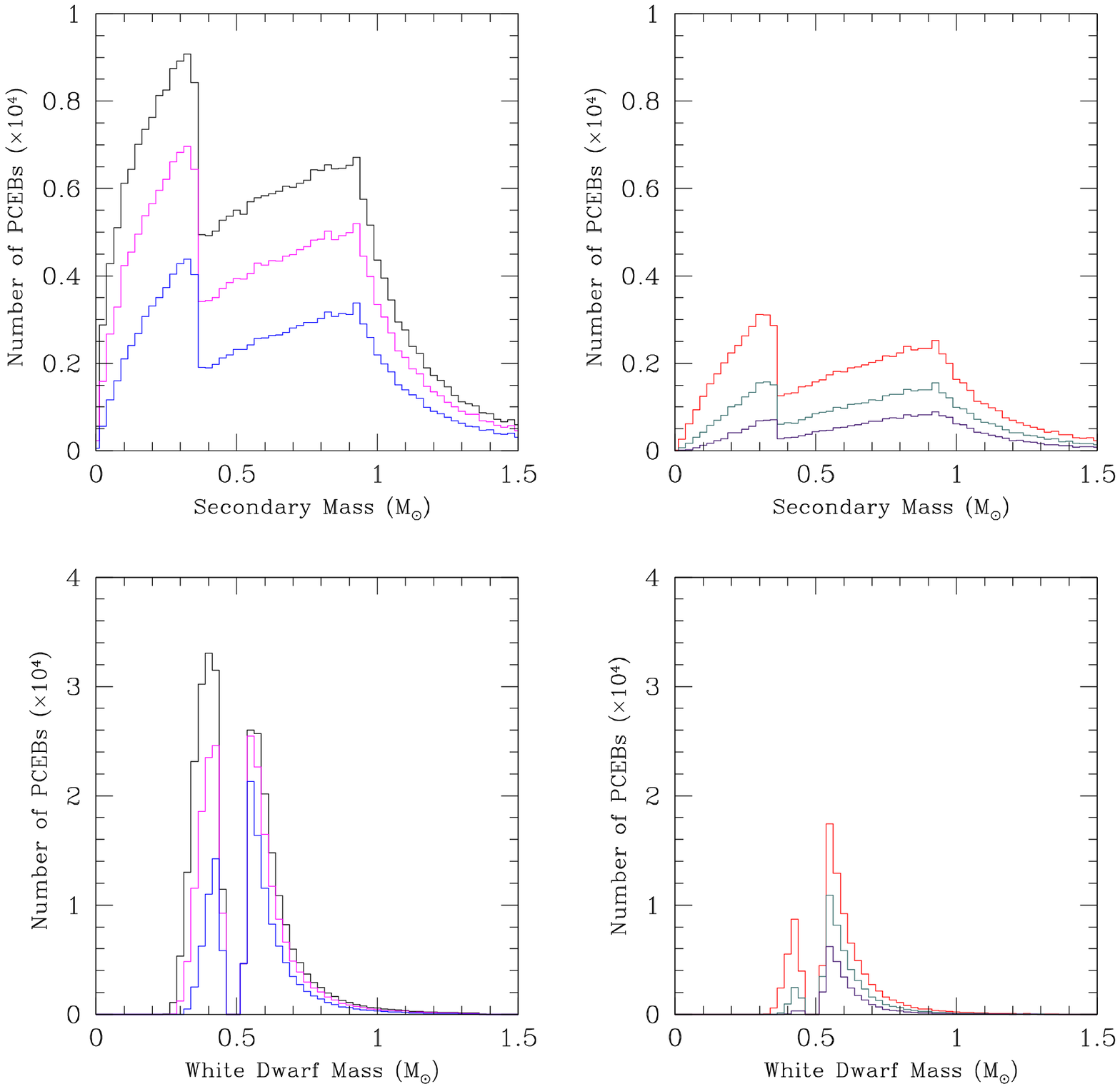}
\figcaption{Theoretical present-day distributions of the secondary masses (top panels) and the WD masses (bottom panels) in PCEBs for $\alpha_{CE}$ = 1.0 (black), 0.6 (magenta), and 0.3 (blue) (left panels) and for $\alpha_{CE}$ = 0.2 (red), 0.1 (green) and 0.05 (purple) (right panels).}
\end{figure}

\begin{figure}
\plotone{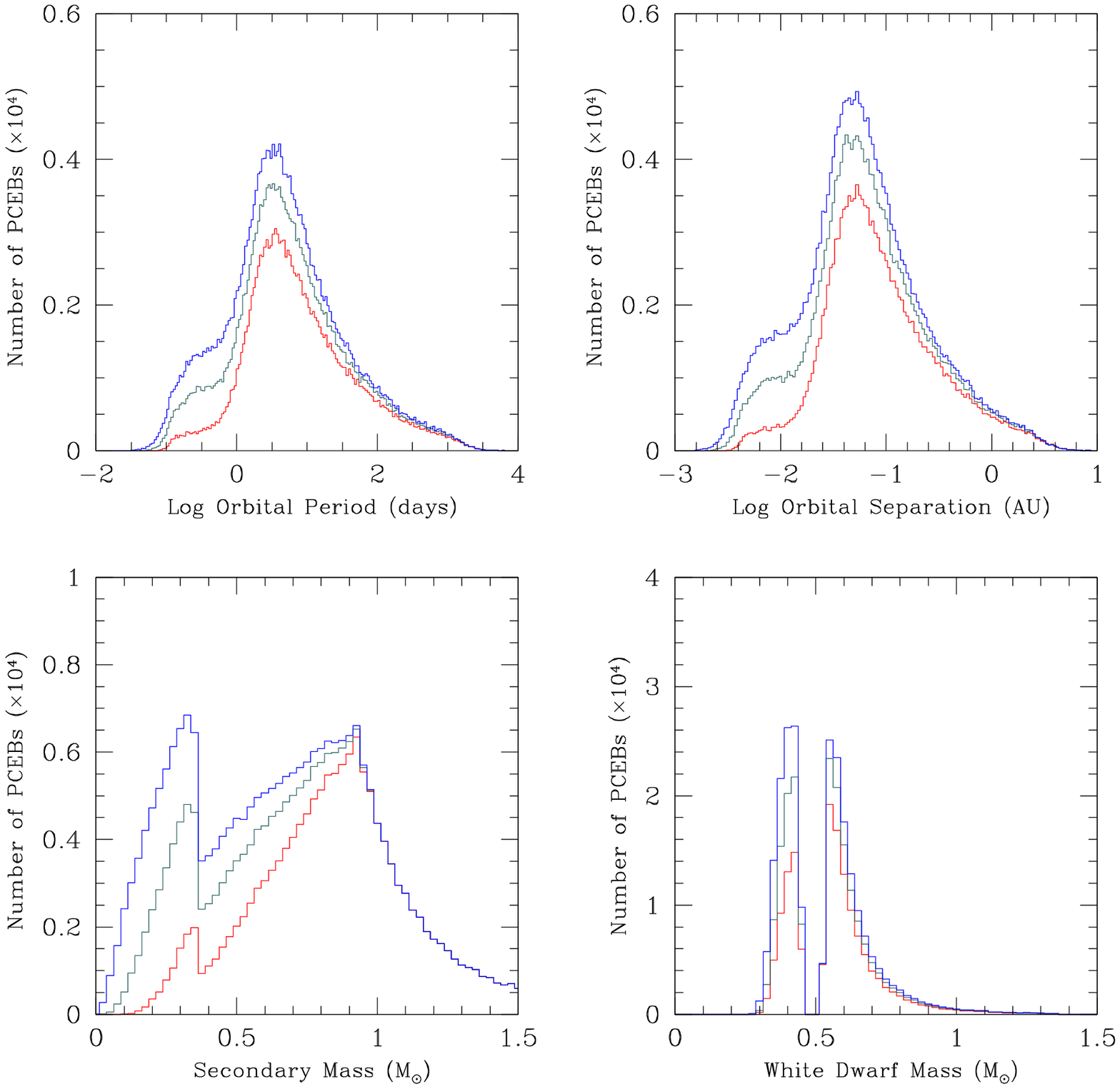}
\figcaption{Theoretical present-day distributions of the orbital periods (top left), the orbital separations (top right), the secondary masses (bottom left) and the WD masses (bottom right) in PCEBs for $\alpha_{CE}$ = $(M_s)^n$, where $n$ = 0.5 (blue), 1.0 (green) and 2.0 (red).}
\end{figure}

\begin{figure}
\plotone{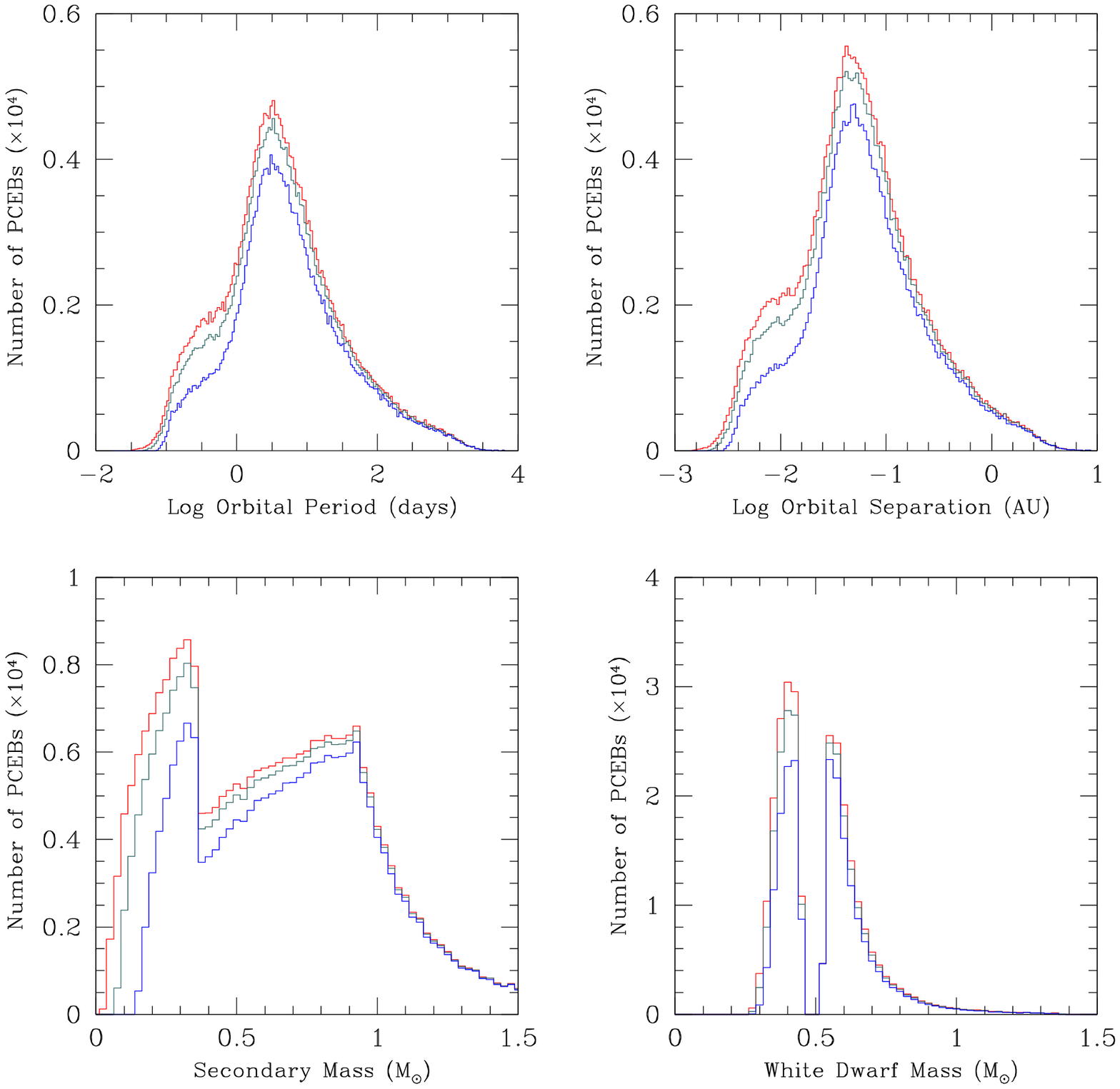}
\figcaption{Theoretical present-day distributions of the orbital periods (top left), the orbital separations (top right), the secondary masses (bottom left) and the WD masses (bottom right) in PCEBs for $\alpha_{CE} = 1-M_{cut}/M_s$, where $M_{cut}$ = 0.0375$\Msun$ (red), 0.075$\Msun$ (green), and 0.15$\Msun$ (blue).}
\end{figure}

\begin{figure}
\plotone{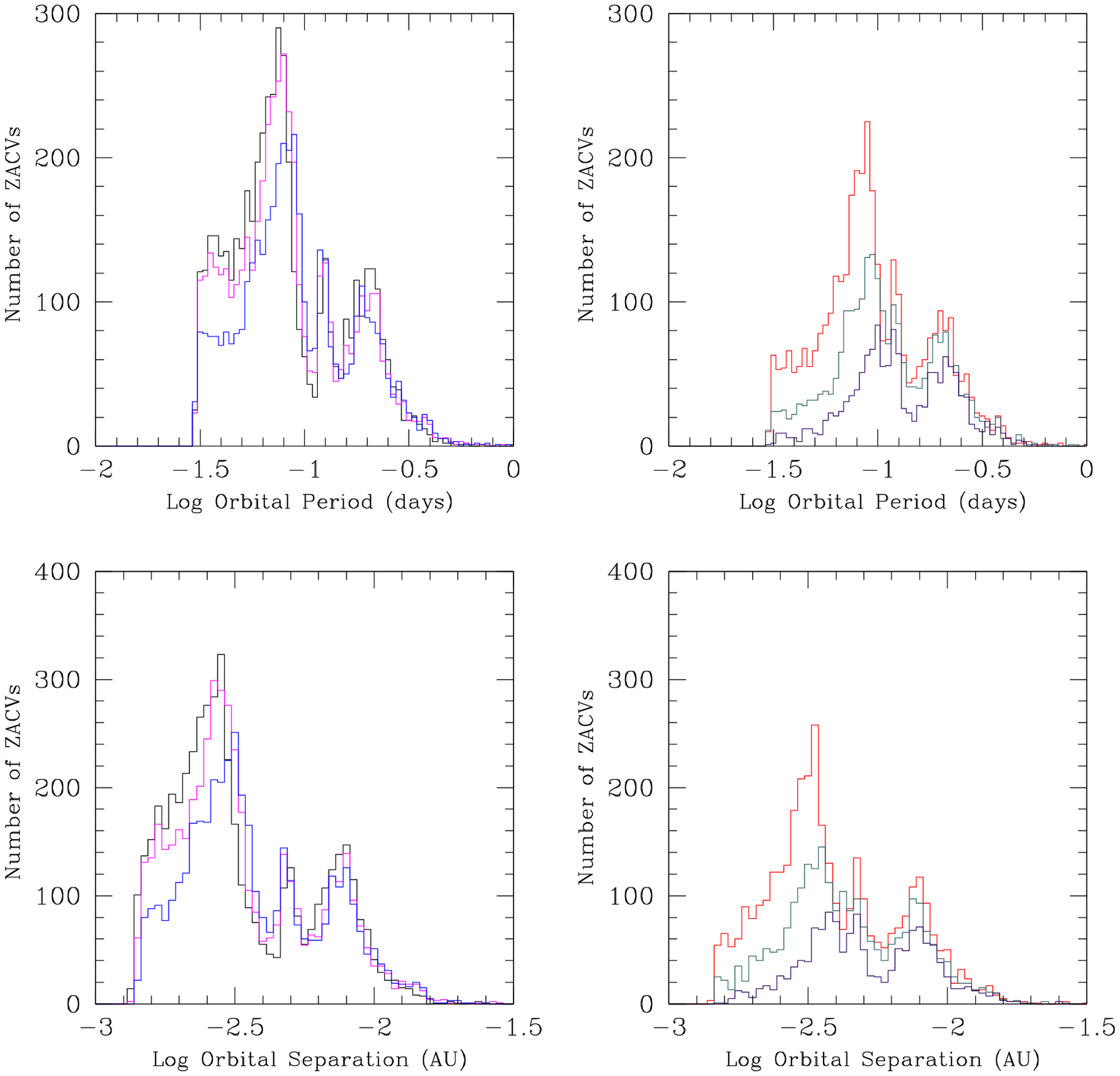}
\figcaption{Theoretical present-day distributions of the orbital periods (top panels) and the orbital separations (bottom panels) in ZACVs for $\alpha_{CE}$ = 1.0 (black), 0.6 (magenta), and 0.3 (blue) (left panels) and for $\alpha_{CE}$ = 0.2 (red), 0.1 (green) and 0.05 (purple) (right panels).}
\end{figure}

\begin{figure}
\plotone{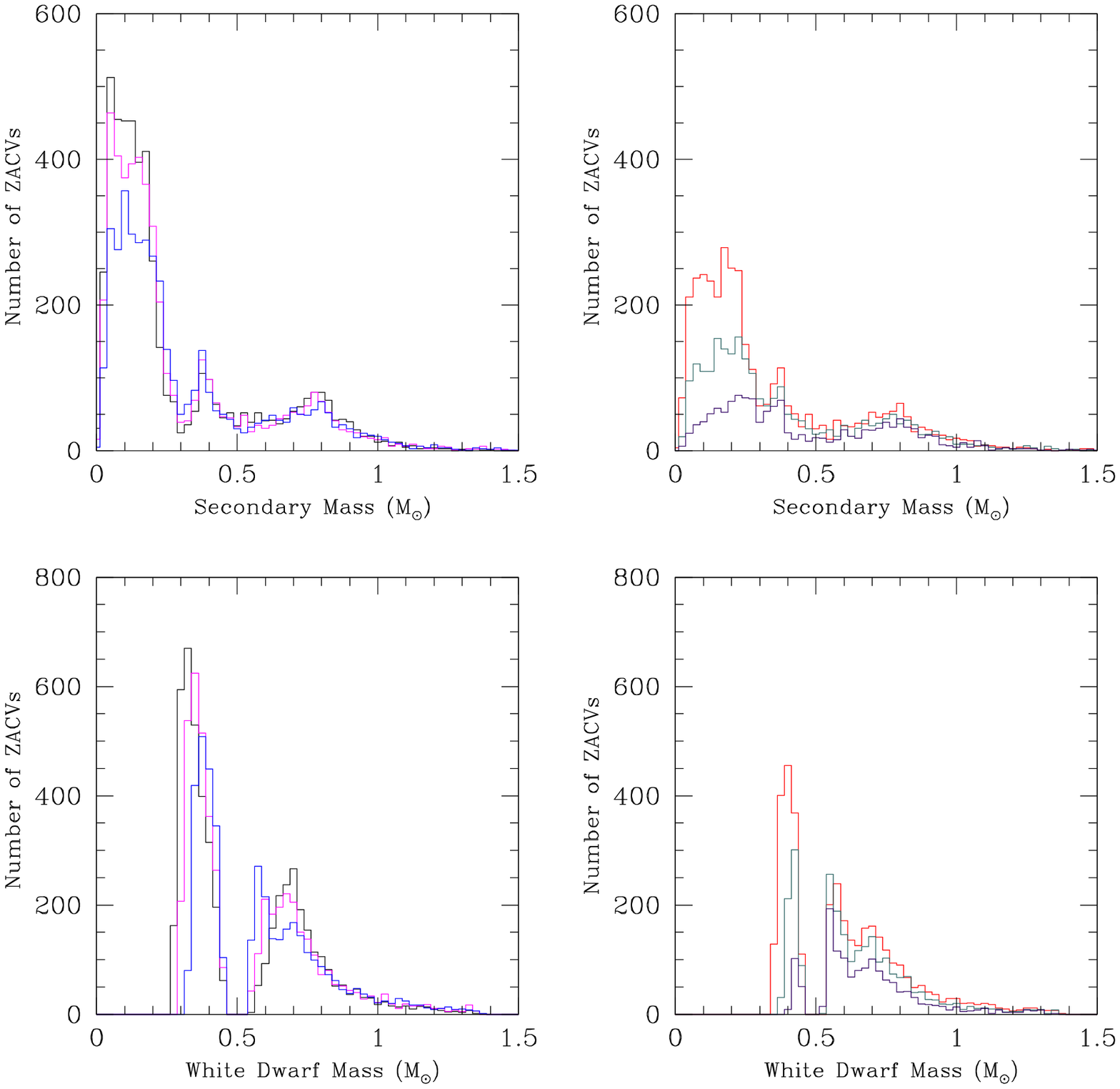}
\figcaption{Theoretical present-day distributions of the secondary masses (top panels) and the WD masses (bottom panels) in ZACVs for $\alpha_{CE}$ = 1.0 (black), 0.6 (magenta), and 0.3 (blue) (left panels) and for $\alpha_{CE}$ = 0.2 (red), 0.1 (green) and 0.05 (purple) (right panels).}
\end{figure}

\begin{figure}
\plotone{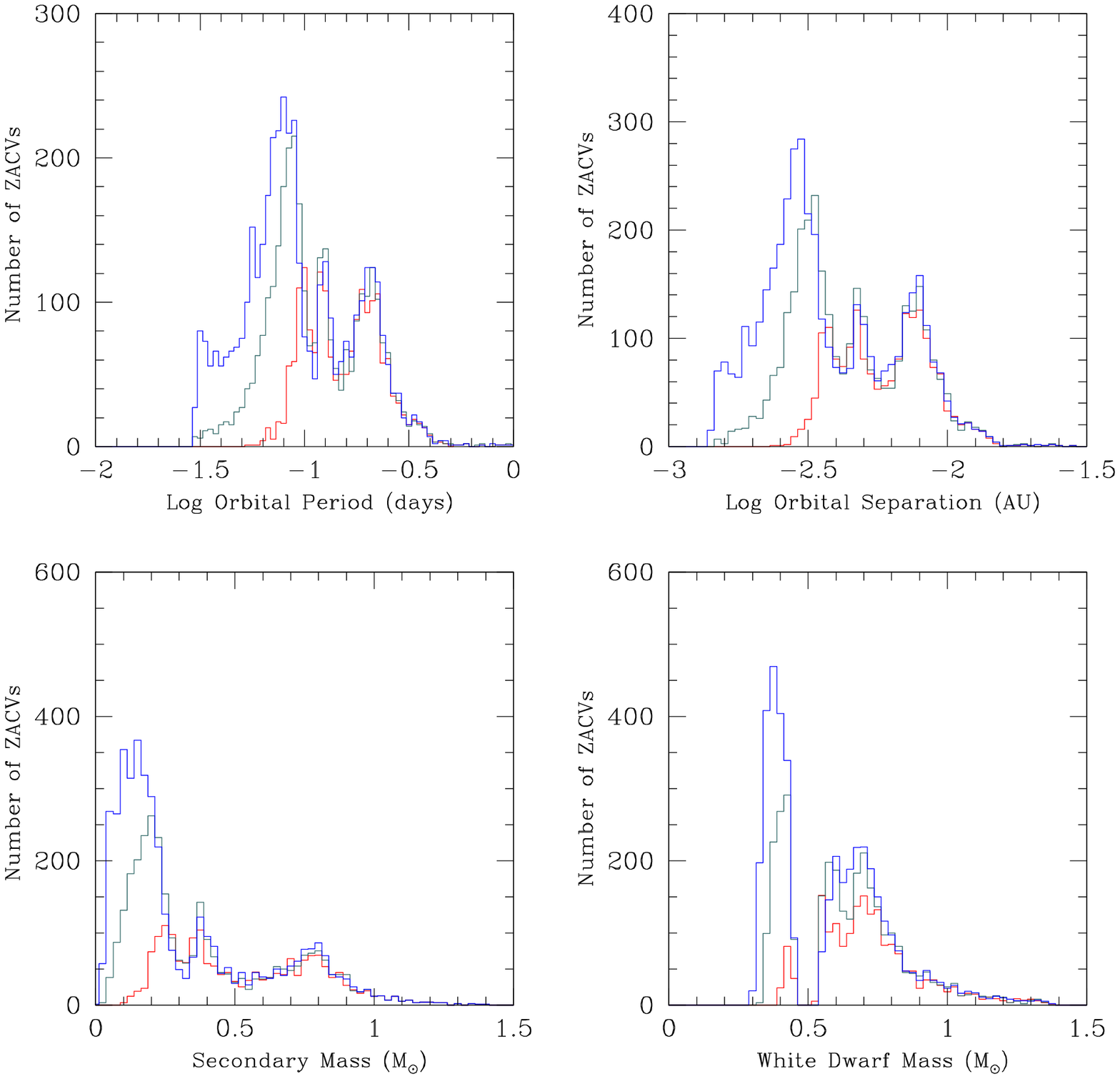}
\figcaption{Theoretical present-day distributions of the orbital periods (top left), the orbital separations (top right), the secondary masses (bottom left) and the WD masses (bottom right) in ZACVs for $\alpha_{CE}$ = $(M_s)^n$, where $n$ = 0.5 (blue), 1.0 (green) and 2.0 (red).}
\end{figure}

\begin{figure}
\plotone{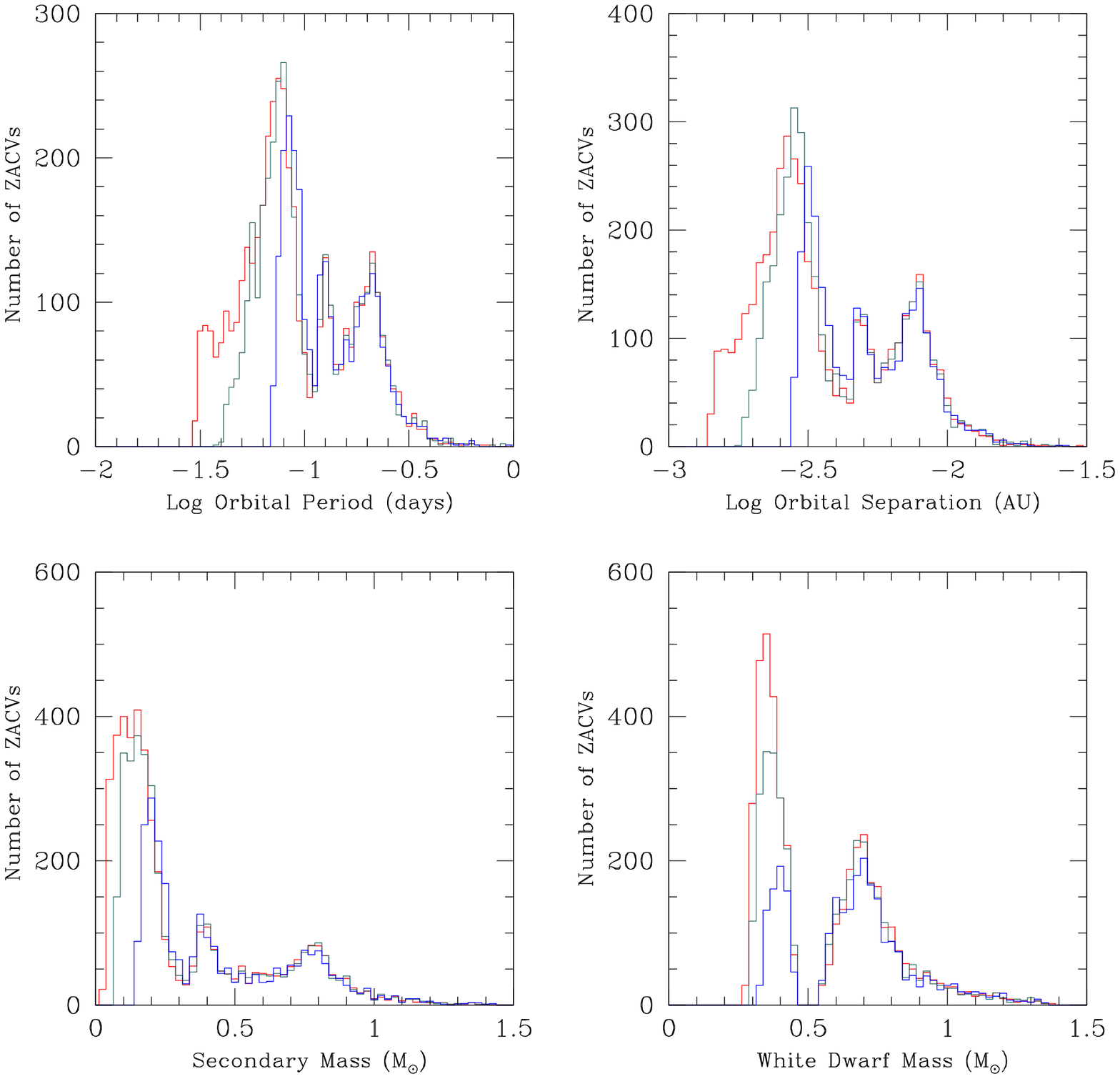}
\figcaption{Theoretical present-day distributions of the orbital periods (top left), the orbital separations (top right), the secondary masses (bottom left) and the WD masses (bottom right) in ZACVs for $\alpha_{CE} = 1-M_{cut}/M_s$, where $M_{cut}$ = 0.0375$\Msun$ (red), 0.075$\Msun$ (green), and 0.15$\Msun$ (blue).}
\end{figure}

\begin{figure}
\plotone{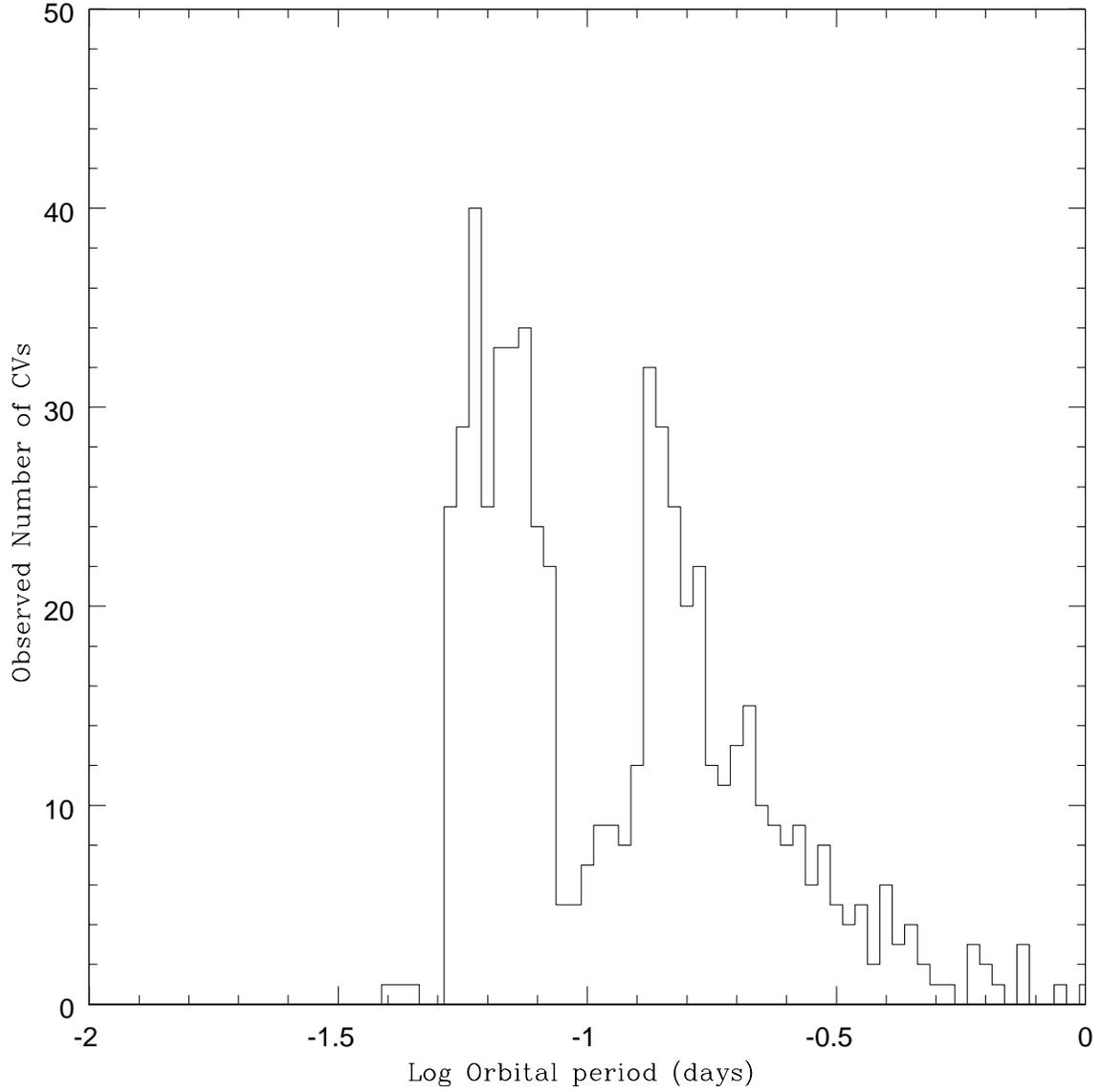}
\figcaption{Observed orbital period distribution in CVs.  The data were taken from the latest online version (RKcat7.6, 1 January 2006) of the \citet{rit03} catalog.  We do not include in the distribution any systems in the catalog that may be possible AM CVn systems.}
\end{figure}

\end{document}